\documentclass[12pt]{article}
\usepackage{amsfonts,amsmath,amssymb}
\usepackage{graphicx}

\newfont{\twelvemsb}{msbm10 scaled\magstep1}
\newfont{\eightmsb}{msbm8}
\newfam\msbfam
\textfont\msbfam=\twelvemsb \scriptfont\msbfam=\eightmsb
\catcode`\@=11
\def\Bbb{\ifmmode\let\next\Bbb@\else
\def\next{\errmessage{Use \string\Bbb\space only in math mode}}\fi\next}
\def\Bbb@#1{{\fam\msbfam{{#1}}}}

\newcommand{\be}{\begin{equation}}
\newcommand{\ee}{\end{equation}}
\newcommand{\ba}{\begin{eqnarray}}
\newcommand{\ea}{\end{eqnarray}}

\begin{document}

\sloppy
\renewcommand{\thefootnote}{\fnsymbol{footnote}}
\newpage
\setcounter{page}{1} \vspace{0.7cm}
\begin{flushright}
\end{flushright}
\vspace*{1cm}
\begin{center}
{\bf The generalised scaling function: a systematic study}\\
\vspace{1.8cm} {\large Davide Fioravanti $^a$, Paolo Grinza $^b$ and
Marco Rossi $^c$
\footnote{E-mail: fioravanti@bo.infn.it, pgrinza@usc.es, rossi@cs.infn.it}}\\
\vspace{.5cm} $^a$ {\em Sezione INFN di Bologna, Dipartimento di Fisica, Universit\`a di Bologna, \\
Via Irnerio 46, Bologna, Italy} \\
\vspace{.3cm} $^b${\em Departamento de Fisica de Particulas,
Universidad de Santiago de Compostela, 15782 Santiago de
Compostela, Spain} \\
\vspace{.3cm} $^c${\em Dipartimento di Fisica dell'Universit\`a
della Calabria and INFN, Gruppo collegato di Cosenza, I-87036
Arcavacata di Rende, Cosenza, Italy}
\end{center}
\renewcommand{\thefootnote}{\arabic{footnote}}
\setcounter{footnote}{0}
\begin{abstract}
{\noindent We describe a procedure for determining the generalised scaling
functions $f_n(g)$ at all the values of the coupling constant. These functions describe
the high spin contribution to the anomalous dimension of large twist operators (in the $sl(2)$ sector) of ${\cal N}=4$ SYM. At fixed $n$, $f_n(g)$ can be obtained by solving a linear integral equation (or, equivalently, a linear system with an infinite number of equations), whose inhomogeneous term only depends on the solutions at smaller $n$. In other words, the solution can be written in a recursive
form and then explicitly worked out in the strong coupling regime.
In this regime, we also emphasise the peculiar convergence of
different quantities ('masses', related to the $f_n(g)$) to the unique mass gap of the $O(6)$
nonlinear sigma model and analyse the first next-to-leading order corrections.}
\end{abstract}
\vspace{6cm}

\newpage

\section{Introduction}

Among the different subsets of local composite operators in planar ${\cal N}=4$ SYM, the $sl(2)$ twist sector has been very much studied under different perspectives and by various means. Under planar limit, {\it i.e.} number of colours $N\rightarrow \infty$ and SYM coupling $g_{YM}\rightarrow 0$,
so that the `t Hooft coupling
\begin{equation}
\lambda =  g_{YM }^2 N= 8 \pi^2 g^2 \,
\end{equation}
stays finite, it is made up of local composite operators with the
form
\begin{equation}
{\mbox {Tr}} ({\cal D}^s {\cal Z}^L)+.... \, , \label {sl2op}
\end{equation}
where ${\cal D}$ is the covariant
derivative acting in all the possible ways on the $L$ complex bosonic fields
${\cal Z}$. The Lorentz spin of these operators is $s$ and $L$ is the $R$-charge which coincides with the twist (classical dimension minus the spin). Moreover, this sector is described --
thanks to the AdS/CFT correspondence \cite{MWGKP} -- by spinning folded closed strings
on $\text{AdS}_5\times\text{S}^5$ spacetime with
$\text{AdS}_5$ and $\text{S}^5$ angular momenta $s$ and $L$, respectively \cite{GKPII, FT}.
In addition, as far as the one loop is concerned, the Bethe Ansatz
problem is equivalent to that of twist operators in QCD
\cite{LIP,BDM}; and this partially justifies the great interest in
the sector (\ref {sl2op}).

As being in a conformal model, suitable superpositions of operators form dilatation operator eigenvectors with definite dimensions (eigenvalues), which are made up of 
a classical part plus an anomalous one. For instance, in the topical sector of operators (\ref{sl2op}) 
this spectral problem shows up dimensions
\begin{equation}
\Delta(g,s,L) = L+s+\gamma (g,s,L) \, , \label{Delta}
\end{equation}
where $\gamma (g,s,L)$ is the anomalous part.
According to the AdS/CFT strong/weak coupling duality, the set of anomalous
dimensions of composite operators in ${\cal N}=4$ SYM coincides with
the energy spectrum of the $\text{AdS}_5\times\text{S}^5$ string theory (\cite{MWGKP,GKPII, FT} and references therein),
although the perturbative regimes are interchanged. The highly
nontrivial problem of evaluating the anomalous part in ${\cal
N}=4$ SYM was greatly simplified by the discovery of
integrability in the purely bosonic $so(6)$ sector at one loop \cite
{MZ}. Later on, this fact has been extended to all the gauge theory sectors and
at all loops in a way which shows up integrability in a weaker sense, but still furnishes the investigators many powerful tools \cite{BS}. More in detail, any
operator ({\it e.g.} of the form (\ref {sl2op})) has been thought of as a state of a 'spin chain', whose hamiltonian is, of course, the dilatation operator itself, although the latter does not have an explicit expression of the spin chain form, but for the first few loops. Nevertheless, the large size ({\it asymptotic}) spectrum has turned out to be exactly described by certain Bethe Ansatz-like equations (the so-called Beisert-Staudacher equations, cf. \cite {BS,BES} and references therein). In other words, the anomalous dimensions coincide with the energies given by the Bethe Ansatz solutions (or roots): this is, of course, a great simplification of the initial spectral problem.    

Unfortunately, this is only part of the full story, albeit the rest should not worry us
in the present context. In fact, an important limitation emerges as a
consequence of the {\it asymptotic} character of the Bethe Ansatz: the latter ought to be  modified by {\it
wrapping effects} as soon as the (site-to-site) interaction range in the loop expansion of the dilatation operator becomes greater than the chain length. In other words, the anomalous dimension given by the asymptotic Bethe Ansatz is {\it in general} correct only up to $L-1$ loops in the (SYM) convergent
perturbative expansion, {\it i.e.} up to the order $g^{2L-2}$. This implies that the asymptotic Bethe Ansatz should give the right result whenever the limit (\ref{jlimit}) below is
applied and the leading contribution (\ref{sud}) considered.

Therefore, let us consider the following large twist and high spin (double
scaling) limit
\begin{equation}
s \rightarrow \infty \, , \quad L \rightarrow \infty \, , \quad
j=\frac {L}{\ln s}={\mbox {fixed}} \, , \label {jlimit}
\end{equation}
in the asymptotic Bethe Ansatz equations describing the $sl(2)$ sector (\ref{sl2op}).
Incidentally, we shall stress how the dual string theory inherits a crucial difference since its semiclassical expansion employs the string tension $\sqrt\lambda\rightarrow +\infty$ as inverse Planck's constant. This means that this limit need to be considered before the scaling (\ref{jlimit})\,(cf. for
instance \cite{FTT} and references therein), thus imposing, at least, a
different limit order with respect to our gauge theory approach: often, instead of $j$, the scaling string variable $\ell \sim j/\sqrt\lambda$ stays naturally fixed (cf.
below for more details). The relevance of this double limit has been
suggested in first instance by \cite{BGK} within the (one-loop) SYM
theory and then motivated in \cite{FTT} and in \cite{AM} within the
string theory dual (see also \cite{GKPII, FT}). In fact, calculations of the latter authors
pointed towards the following generalisation (at all loops) for
the anomalous dimension formula found in \cite{BGK}
\begin{equation}
\gamma (g,s,L)=f(g,j)\ln s + \dots \,\, . \label{sud}
\end{equation}
Moreover, by describing the Bethe Ansatz energy through a non-linear
integral equation (like in other integrable theories \cite{FMQR}),
this Sudakov scaling has been remarkably confirmed in \cite{FRS}.
There this statement was argued by computing iteratively
the solution of a (inhomogeneous) linear integral equation (Neumann expansion)
and then, thereof, {\it the generalised scaling function}, $f(g,j)$ at the first orders in $j$
and $g^2$: more precisely the first orders in $g^2$ have been
computed for the first {\it generalised scaling functions} $f_n(g)$,
forming the crucial expansion (see below for the motivation)
\begin{equation}
f(g,j)=\sum _{n=0}^{\infty} f_n(g)j^n \, .
\label{j=0exp}
\end{equation}
As a by-product, the reasonable conjecture has been put forward that
the two-variable function $f(g,j)$ should be bi-analytic around zero (in $g$ for
fixed $j$ and in $j$ for fixed $g$). Of course, a reliable test of the AdS/CFT correspondence
requires the knowledge of the  $f_n(g)$ also for large values of the coupling $g$, as a consequence of the semiclassical nature of string expansion. This fact has been recently experienced in a particular, but peculiar case, namely the (large $g$) asymptotic expansion of $f_0(g)=f(g)$  and the comparison with string theory results (\cite{BBKS, FTT, BKK,KSV, CK} and references therein).

In this context, in paper \cite{BFR2} we have studied the large $s$ limit at finite $L$, showing
how to obtain the contributions beyond the leading scaling function $f(g)$, by means of one linear integral equation, which does not differ from the so-called BES equation (which covers the case $j=0$, cf. the second one of \cite{BES}), but for the inhomogeneous term. In this respect, our approach was different with respect to that of \cite {FRS}, as the latter needs to take into account also non-linear terms in the integral equation and anomalous dimension expression (cf. also below for other details).

Albeit important in itself, this first step is also important for the study of the large $L$ limit (\ref {jlimit}), which is indeed the main aim of the present systematic study. Actually, a suitable
modification of the LIE of \cite{BFR2} has been already exploited and explored in
\cite{FGR} to derive, in the scaling (\ref{jlimit}), still a LIE, namely (\ref {sigmaeq2}) below. This equation yields the same leading scaling function $f(g,j)$ (of the expansion (\ref{sud})) as the LIE in \cite{FRS}. Yet, we will argue in the next section how our LIE should also predict the form of the dots in (\ref{sud}). More precisely, we would expect a $O((\ln s)^0)$ ($j$-dependent) correction to the leading Sudakov scaling, {\it i.e.}
\begin{equation}
\gamma (g,s,L)=f(g,j)\ln s + f^{(0)}(g,j) +  \dots \, \, .
\label{corrsud}
\end{equation}
Furthermore, the same linear integral equation (\ref{sigmaeq2}) still controls this next-to-leading order (nlo), $f^{(0)}(g,j)$. Now, similarly, we may imagine that the dots should initially be inverse integer powers of $\ln s$, with coefficients, at each power, depending on $g$ and $j$. Afterwards, inverse integer powers of $s$ should also enter the stage, but they are determined by the complete non-linear integral equation (NLIE) of \cite{BFR2} \footnote{These are also corrected by wrapping effects.}. However, in this paper we will constrain ourselves to the leading Sudakov factor $f(g,j)$, leaving the analysis of its corrections for future publications.

Actually, in \cite {FGR} we have initiated the study of the strong coupling regime of the first generalised scaling function $f_1(g)$ and have shown the proportionality of its leading order to the
mass gap $m(g)$ (see (\ref{mgap}) below) of the $O(6)$ nonlinear sigma
model (NLSM). This gives a first positive test, in the strong coupling regime $j\ll m(g)$ of the NLSM, for the Alday-Maldacena proposal \cite{AM}. This claims that as long as $g \gg j$ the quantity $f(g,j)+j$ should coincide with the $O(6)$ NLSM energy density. The latter was expanded and checked for the first orders in the perturbative regime $j\gg m(g)$ of the NLSM by \cite{AM}. Hence, our test was a first indication in another valuable region of the NLSM, {\it i.e.} $j\ll m(g)$, where the free energy series is, besides, convergent \cite{BF}. Afterwards, the embedding of the
$O(6)$ NLSM into ${\cal N}=4$ SYM at large $g$ was brilliantly shown in a formal
way by \cite {BK}, where the leading strong coupling contribution of $f_3(g)$ was computed too. In a
contemporaneous paper \cite{FGR2}, starting from the our linear integral equation \cite{FGR}, we have set down the initial ideas for a systematic study of all the $f_n(g)$ and confined our study to the first four $f_1(g)$, $f_2(g)$, $f_3(g)$ and $f_4(g)$, by finding for them some analytic relations and expressions. These have been then evaluated numerically with additional analytic results for large $g$, finding agreement with the suitable results  from the $O(6)$ NLSM \cite{BF}.  Furthermore, the agreement on $f_4(g)$ is highly nontrivial, since it contains the details of the specific interaction in the $O(6)$ NLSM. For completeness sake, all these results will be reported in the following as well.

In the present paper we want to present a systematic approach to the computation of all the generalised scaling functions $f_n(g)$ and, consequently, of $f(g,j)$ according to the expansion about $j=0$ (\ref{j=0exp}). In section 2 we will write a linear integral equation for the (higher loop) root/hole density which describes the anomalous dimension via a linear integral in the limit (\ref{jlimit}). Then, the problem of computing the generalised scaling function $f(g,j)$ (\ref{sud}) will be achieved by expanding the density around $j=0$, analogously to (\ref{j=0exp}). The $n$-th coefficient ($n$-th 'density') of this expansion gives the $n$-th generalised scaling function $f_n(g)$ via a Kotikov-Lipatov-like \cite{KL} formula (\ref {egs2}) and satisfies an integral equation (of Fredholm type) whose inhomogeneous term involves specific values of the $m$-th densities with $m\leq n-3$: this fact clearly permits a recursive solution. In section 3 the general $n$-th integral equation will be re-written as a linear system for an infinite dimensional vector, whose first component is simply proportional to $f_n(g)$. In section 4 we will make the recursive procedure more explicit and write down systematically the solution of the $n$-th system in terms of the solutions of simpler systems \footnote{Because of their 'simplicity', which expresses itself mainly through the possibility of writing their solution in terms of the BES solution \cite{BES, BKK}, we will call them 'reduced' systems.} and of the values of the root/hole density and its derivatives
in zero. In section 5 we study extensively the strong coupling limit, $g \rightarrow +\infty$. First, we write down the asymptotic (power-like) expressions for the root/hole density and its derivatives. Then, we set down a recursive method for computing the non-analytic correction to them and to the various scaling functions $f_n(g)$. As an example we make this procedure explicit for the first cases $f_3(g), \ldots , f_8(g)$ (subsection 5.1) and then show that these all tend to their $O(6)$ nonlinear sigma model (NLSM) prediction (given in terms of the mass-gap). Finally, a quantitative discussion on the correction to this limit is presented (subsection 5.2). Some perspectives and conclusions are presented in section 6.

\section{High spin equations}
\setcounter{equation}{0}

In the framework of integrability in ${\cal N}=4$ SYM, it was useful
\cite{NOI} to rewrite the Bethe equations as non-linear integral
equations \cite {FMQR}.
In particular, this approach was pursued for
the $sl(2)$ sector of the theory (see \cite {FRS,BFR2}), since
it allows to evaluate in a rigorous way all the subleading terms in the high spin expansion.
In fact, as long as the leading $O(\ln s )$ term is under investigation, a simpler derivation of the relevant
linear equations based on the density of roots can be used: this was the way followed in
the seminal papers on the ES and BES equations (first and second of \cite {BES}, respectively). Nevertheless, once the subsequent $\ln s$ orders enter the stage  -- as for the generalised scaling function $f(g,j)$ --, it is unclear to which extent a linear equation for the density may be correctly derived. Actually, this is part of the aims of the non-linear integral equation method, namely to reproduce (rigorously the solution to) the density equation as leading large volume contribution, by taking under control the non-linear terms \cite{FMQR}. In this spirit, \cite{FRS} has improved the analysis in \cite{BES} by evaluating at which order the non-linear terms would have contributed; as a consequence, a linear integral equation describing $f(g,j)$ has been derived. However, we will start from the non-linear integral equation derived in \cite {BFR2}, since non-linearity starts contributing at larger order ($O((\ln s )^{-1})$, see discussion after equation (\ref {normale})), thus making possible the study of the first subleading correction in future studies. Of course, for the further corrections the full non-linear integral equation will be crucial.

In the $sl(2)$ sector states of twist $L$ are
described by $s$ Bethe roots, which localize in an interval $[-b,b]$
of the real line, and $L$ 'holes' \cite{BGK, BES, FRS, BFR2}. For any state, two holes lie
outside the interval $[-b,b]$ and the remaining $L-2$ holes lie
inside this interval.
For the lowest anomalous dimension state
(ground state) -- which is the state we are interested in -- the
$(L-2)$ internal holes localise in the interval
$[-c,c]$, $c<b$, and in this interval no roots are
present\footnote{The existence of a 'separator', $c$, between roots and holes is a non-obvious and technically important issue.}. The non-linear integral equation for states of the $sl(2)$ sector
involves two functions $F(u)$ and $G(u,v)$ satisfying linear
integral equations \cite{BFR2}. It is convenient to split $F(u)$ into its one-loop
$F_0(u)$ and higher than one loop $F^H(u)$ contributions and to
define the functions $\sigma _H(u) =\frac {d}{du} F^H(u)$ and
$\sigma _0(u)=\frac {d}{du} F_0(u)$. In the limit (\ref {jlimit})
({\it i.e.} $L\rightarrow +\infty$ with $j$ fixed) these functions (depend on $j$ and)
acquire the meaning of, respectively, higher than one loop and one
loop density of both Bethe roots and holes. At the leading order
$O(\ln s )$ they satisfy linear integral equations
below (\ref {hatsigma0}), (\ref {sigmaeq2}), respectively.

Regarding the ground state, in the scaling (\ref {jlimit}) the densities determine the anomalous
dimension in a simple form \footnote {The
parameter $b_0>c$ is the one loop contribution to $b$: therefore, it
depends on $s$ through the solution of the linear equation for
$F_0(u)$ (see \cite {BFR2}).},
\begin{eqnarray}
\gamma (g,s,L)&=&-g^2\int _{-b_0}^{b_0}\frac {dv}{2\pi} \left [\frac
{i}{x^+(v)}-\frac {i}{x^-(v)}\right ] \sigma_0(v)+
\nonumber \\
&+&g^2 \int _{-\infty}^{+\infty} \frac {dv}{2\pi} \chi _c(v) \left [
\frac {i}{x^+(v)}-\frac {i}{x^-(v)}\right ] [  \sigma_0(v)+
\sigma _H(v) ] -  \label{egs} \\
&-& g^2\int _{-\infty}^{+\infty}\frac {dv}{2\pi} \left [\frac
{i}{x^+(v)}-\frac {i}{x^-(v)}\right ] \sigma_H(v) + O\left(({\ln
s})^{-1}\right ) \, , \nonumber
\end{eqnarray}
where we introduced the function $\chi  _{c} (u)$ which equals $1$
if $-c\leq u \leq c$, where the internal holes concentrate, and $0$
otherwise. This
means that, as far as the computation of the generalised scaling
functions is concerned, one can rely on (\ref {egs}). The terms
depending on $\sigma _0(u)$ get more manageable after using the
important relation
\begin{equation}
\int _{-b_0}^{b_0} dv f(v) \sigma _0(v)= \int _{-\infty}^{+\infty}
dv f(v) \sigma _0^s(v) +O\left (({\ln s})^{-1}\right ) \, , \label{1loop}
\end{equation}
where the Fourier transform of the function $\sigma _0^s(v)$
satisfies the integral equation,
\begin{equation}
\hat \sigma _0^s(k)=-4\pi \frac {\frac {L}{2}-e^{-\frac {|k|}{2}}
\cos \frac {ks}{\sqrt {2}}} {2\sinh \frac {|k|}{2}}-\frac {e^{-\frac
{|k|}{2}}}{2\sinh \frac {|k|}{2}} \int  _{-\infty}^{+\infty} du
e^{iku} \chi _{c_0} (u) \sigma _0^s(u)-4\pi \delta (k) \ln 2 \, ,
\label {sigma0}
\end{equation}
with the parameter $c_0$ such that the normalization
condition\footnote {The physical meaning of (\ref {sigma0cond}) is
that, for $-c_0\leq u\leq c_0$, $\sigma _0^s (u)$ approximates the
density of holes, that in the one loop theory fill the interval
$[-c_0, c_0]$, where no roots are present.}
\begin{equation}
\int  _{-\infty }^{+\infty} du \chi _{c_0} (u)\sigma
_0^s(u)=-2\pi(L-2) + O\left(({\ln s})^{-1}\right ) \,  \label {sigma0cond}
\end{equation}
holds.
Formula (\ref {1loop}) was introduced in \cite {BFR2} (it is formula
(3.52) there), upon taking inspiration from analogous simplifications
used in the ES paper (first reference of \cite{BES}), though for the more than one loop density.
In the double limit (\ref {jlimit}), {\it i.e.} considering
$j$ fixed, the above $j$-depending remainders are $O\left( ({\ln s})^{-1}\right )$ and are given by
non-linear terms we neglected when writing previous equations. This means that the linearity of equations extends
also to the subsequent order $O\left((\ln s )^0\right)$, and thus eventually to $f^{(0)}(g,j)$ of (\ref{corrsud}) : this case will be object of future investigations and publications. In this paper we will constrain ourselves to the leading order
$O(\ln s )$. Therefore, we can neglect the $\delta$-term in (\ref {sigma0})
and we are left with the following equations, describing the one loop theory:
\begin{eqnarray}
\hat \sigma _0^s(k)&=&-4\pi \frac {\frac {L}{2}-e^{-\frac {|k|}{2}}
\cos \frac {ks}{\sqrt {2}}}
{2\sinh \frac {|k|}{2}}-\nonumber \\
&-& \frac {e^{-\frac {|k|}{2}}}{\sinh \frac {|k|}{2}} \int
_{-\infty}^{+\infty} \frac {dh}{2\pi} \hat \sigma _0^s(h) \frac
{\sin (k-h)c_0}{k-h}
 \, , \label {hatsigma0} \\
2 \int  _{-\infty }^{+\infty} \frac {dk}{2\pi} \hat \sigma _0^s(k)
\frac {\sin kc_0}{k} &=&-2\pi(L-2) \, .
\label {hatsigma0cond}
\end{eqnarray}
These relations have to be solved together and give, for any values
of $L$, $c_0$ and $\hat \sigma _0^s(k)$ at the leading order $O(\ln s)$. In
the limit (\ref {jlimit})
$\sigma _0^s(u)$ and $c_0$ expand as,
\begin{equation}
\sigma _0^s(u)=[\sum _{n=0}^{\infty} {\sigma _0^s}^{(n)}(u)j^n]\ln s
+\ldots  \, , \quad c_0=\sum _{n=1}^{\infty}
c_0^{(n)}j^n + \ldots \, , \label {c0series}
\end{equation}
where dots stand for subleading corrections,
and it is not difficult to give, see for instance \cite {FGR2}, the values of the
first two coefficients of the expansion of $c_0$:
\begin{equation}
c_0^{(1)}=\frac {\pi}{4} \, , \quad c_0^{(2)}=-\frac {\pi}{4}\ln 2
\, .
\end{equation}
In order to study the higher than one loop density in the limit (\ref {jlimit}),
we start from (4.10) of \cite {BFR2}. We remove the third, the fourth, the fifth, the seventh and the eighth term in the right hand side of that equation, since they are all $O(1/s)$. Moreover, using the
localization \cite{BES} of the higher than one loop density, in all the
integrations involving $\frac {d}{dv} F_H(v)=\sigma _H(v)+O(1/s)$ we replace
the extremes $\pm b$ with $\pm \infty$. On the other hand, in the integrations involving $\frac {d}{dv}
F_0(v)=\sigma _0(v)+O(1/s)$ we can replace $b$ with $b_0$ and then use (\ref {1loop}).
Finally, we replace the sums over internal holes with integrals involving the density.
After doing all these manipulations, we obtain that the higher than one loop density satisfies the
linear integral equation\footnote {The quantity $\theta (u,v)$
appearing in (\ref {sigmaeq}) is the well known dressing factor. For
related notations, we refer to the second reference of \cite {BES}.}, at the leading order
$O(\ln s)$,
\begin{eqnarray}
\sigma _{H}(u)&=& -iL \frac {d}{du} \ln \left ( \frac {1+\frac {g^2}
{2{x^-(u)}^2}}{1+\frac {g^2}{2{x^+(u)}^2}} \right ) + \frac {i}{\pi}
\int _{-\infty}^{+\infty} dv  \chi _{c} (v)\Bigl [\frac {d}{du} \ln
\left ( \frac {1-\frac {g^2}{2x^+(u)x^-(v)} }{1-\frac
{g^2}{2x^-(u)x^+(v)}} \right )
+ \nonumber \\
&+& i \frac {d}{du}\theta (u,v)  +i \frac {1}{1+(u-v)^2}\Bigr
][\sigma _0^s(v) +\sigma _H(v)]
\nonumber \\
&-&  \frac {i}{\pi} \int _{-\infty}^{+\infty} dv \frac {d}{du}\left
[ \ln \left ( \frac {1-\frac {g^2}{2x^+(u)x^-(v)} }{1-\frac
{g^2}{2x^-(u)x^+(v)}} \right )
+i \theta (u,v) \right ] \sigma _0^s(v)+  \label {sigmaeq} \\
&+&\int _{-\infty}^{+\infty} \frac {dv}{\pi} \frac {1}{1+(u-v)^2}\sigma _{H}(v) +\int _{-\infty}^{+\infty} \frac {dv}{\pi} \chi _{c_0} (v) \frac {1}{1+(u-v)^2}\sigma _{0}^s(v)-\nonumber  \\
&-& \frac {i}{\pi} \int _{-\infty}^{+\infty} dv \frac {d}{du} \left
[ \ln \left ( \frac {1-\frac {g^2}{2x^+(u)x^-(v)} }{1- \frac
{g^2}{2x^-(u)x^+(v)}} \right )+i \theta (u,v) \right ]  \sigma
_{H}(v) \, , \nonumber
\end{eqnarray}
which has to be solved knowing (\ref {hatsigma0}) and together with the conditions,
\begin{eqnarray}
&& \int  _{-\infty}^{+\infty} du \chi  _c (u) [\sigma _0^s(u)+\sigma
_{H}(u)] =-2\pi(L-2)  \, , \nonumber \\
&& \label {normliz} \\
&& \int  _{-\infty }^{+\infty} du \chi _{c_0} (u)\sigma
_0^s(u)=-2\pi(L-2)  \, . \nonumber
\end{eqnarray}
As in the one loop case, it is convenient to rewrite, in terms of
Fourier transforms, equation (\ref {sigmaeq}),
\begin{eqnarray}
&&\hat \sigma _H(k)=\pi L \frac {1-J_0({\sqrt {2}}gk)}{\sinh \frac
{|k|}{2}}
+  \nonumber \\
&+& \frac {1}{\sinh \frac {|k|}{2}} \int _{-\infty }^{+\infty} \frac
{dh}{|h|} \Bigl [ \sum _{r=1}^{\infty} r (-1)^{r+1}J_r({\sqrt
{2}}gk) J_r({\sqrt {2}}gh)\frac {1-{\mbox {sgn}}(kh)}{2}
e^{-\frac {|h|}{2}} + \nonumber \\
&+&{\mbox {sgn}} (h) \sum _{r=2}^{\infty}\sum _{\nu =0}^{\infty}
c_{r,r+1+2\nu}(g)(-1)^{r+\nu}e^{-\frac {|h|}{2}} \Bigl (
J_{r-1}({\sqrt {2}}gk) J_{r+2\nu}({\sqrt {2}}gh)-  \label {sigmaeq2} \\
&-& J_{r-1}({\sqrt {2}}gh) J_{r+2\nu}({\sqrt {2}}gk)\Bigr ) \Bigr ]
\Bigl [\hat \sigma _0^s(h)+ \hat \sigma _H(h)-\int  _{-\infty
}^{+\infty} \frac {dp}{2\pi} \left (\hat \sigma _0^s(p)+\hat \sigma
_H(p) \right ) 2 \frac {\sin (h-p)c} {h-p} \Bigr]
- \nonumber \\
&-&   \frac {e^{-\frac {|k|}{2}}}{\sinh \frac {|k|}{2}} \int
_{-\infty }^{+\infty} \frac {dp}{2\pi} \left (\hat \sigma
_0^s(p)+\hat \sigma _H(p) \right ) \frac {\sin (k-p)c} {k-p} + \frac
{e^{-\frac {|k|}{2}}}{\sinh \frac {|k|}{2}} \int  _{-\infty
}^{+\infty} \frac {dp}{2\pi}
\hat \sigma _0^s(p) \frac {\sin (k-p)c_0} {k-p}  \, , \nonumber
\end{eqnarray}
and also the normalization conditions,
\begin{eqnarray}
2 \int  _{-\infty }^{+\infty} \frac {dk}{2\pi} \hat \sigma _0^s(k)
\frac {\sin kc_0}{k} &=&-2\pi(L-2)  \, , \nonumber \\
\label {normale} \\
2 \int  _{-\infty }^{+\infty} \frac {dk}{2\pi} [\hat \sigma
_0^s(k)+\hat \sigma _H(k)] \frac {\sin kc}{k} &=&-2\pi(L-2) \, . \nonumber
\end{eqnarray}
We now briefly comment on equations (\ref {sigmaeq2}, \ref {normale}).
If such equations are supplemented with equation (\ref {hatsigma0}) for the one loop density,
they are sufficiently precise to capture the generalised scaling function $f(g,j)$
appearing in the expansion (\ref {corrsud}). However, if in the one loop
density equation one includes the $\delta$-term, {\it i.e.} if one uses (\ref {sigma0}),
linear equations (\ref {sigmaeq2}, \ref {normale}) are able to capture
also the next-to-leading correction $f^{(0)}(g,j)$ in the scaling
(\ref {corrsud}). This marks a difference with linear equation
of \cite {FRS}, which apparently seems (to us) not able to give next-to-leading
corrections to $f(g,j)$.

Again, in the limit (\ref {jlimit}) both $c$ and $\sigma _H(u)$
expand in powers of $j$,
\begin{equation}
\sigma _H(u)=[\sum _{n=0}^{\infty} \sigma _H^{(n)}(u)j^n]\ln s
+\ldots \, , \quad c=\sum _{n=1}^{\infty} c^{(n)}j^n + \ldots
\, , \label {csigmaexp}
\end{equation}
so that we find convenient to use the all loops density\footnote
{The one loop quantity $\sigma _0^s (u)$ is an approximation,
according to (\ref {1loop}), to the real one loop density. In the limit
(\ref {jlimit})
such approximation is completely justified.}
\begin{equation}
\sigma (u)=\sigma _H(u)+\sigma _0^s (u) \, , \label {allloopsdens}
\end{equation}
which expands as
\begin{equation}
\sigma (u)=[\sum _{n=0}^{\infty} \sigma ^{(n)}(u)j^n]\ln s +\ldots  \, . \quad
\end{equation}
As in the one loop case, it is easy to give (see \cite {FGR2}) the first orders in the
expansion of $c$ in the limit (\ref {jlimit}):
\begin{equation}
c^{(1)}=\frac {\pi}{4-\sigma _H^{(0)}(0)} \, , \quad c^{(2)}=-\pi
\frac {4\ln 2 -\sigma _H^{(1)}(0)}{[4-\sigma _H^{(0)}(0)]^2} \, .
\end{equation}
Finally, by comparing (\ref {egs}) with (\ref {sigmaeq2}), we
can generalise the Kotikov-Lipatov relation (concerning only the cusp anomalous dimension) \cite {KL}:
\begin{equation}
\gamma (g,s,L)=\frac {1}{\pi} \lim _{k\rightarrow 0}\hat \sigma
_H(k)  \label {egs2}\, .
\end{equation}
This equality implies, very simply, that
\begin{equation}
f_n(g)=\frac {1}{\pi} \hat \sigma _H^{(n)}(0) \, . \label {fnsigman}
\end{equation}
A systematic approach to the computation of $f_n(g)$ using the
solution of (\ref {sigmaeq2}) and explicit formul{\ae} which allow
their exact determination at strong coupling is the main issue and
the main result of this paper.

\section{On the calculation of the generalized scaling functions}

From result (\ref {fnsigman}) we realize that the generalised
scaling functions $f_n(g)$ can be extracted from the $n$-th
component $\sigma _H^{(n)}(u)$ of the solution of (\ref {sigmaeq})
in the limit (\ref {jlimit}). We are therefore going to analyze in a
systematic way (\ref {sigmaeq}) or (\ref {sigmaeq2}) in such a
limit. Equations for $\sigma _H^{(0)}(u)$ and $\sigma _H^{(1)}(u)$
were already studied: the former is the well known BES equation
(second reference of \cite {BES}), the latter was treated in detail in \cite
{FGR}. Since in this paper we will use results involving $\sigma
_H^{(0)}(u)$ and $\sigma _H^{(1)}(u)$, we will treat also briefly
these two cases, which, in addition, need to be considered
separately from the rest of the $\sigma _H^{(n)}(u)$.

\subsection{The BES equation}
\setcounter{equation}{0}

If we restrict (\ref {sigmaeq2}) to the component proportional to
$\ln s \cdot j^0$ we obtain, of course, the BES equation for $\hat \sigma _H^{(0)}(k)$. We
now briefly describe - using results of the first of \cite {BBKS} - how to rewrite
such equation in form of an infinite system, suitable for our future manipulations.
We first define
\begin{equation}
S^{(0)}(k)= \frac {2\sinh \frac {|k|}{2}}{2\pi  |k|}\hat \sigma
_H^{(0)}(k) \label {S0def} \, ,
\end{equation}
and then expand $S^{(0)}(k)$, $k\geq 0$, in series of Bessel
functions,
\begin{equation}
S^{(0)}(k)=\sum _{p=1}^{\infty}S^{(0)}_{2p}(g)\frac {J_{2p}({\sqrt
{2}}gk)}{k}+\sum _{p=1}^{\infty}S^{(0)}_{2p-1}(g)\frac
{J_{2p-1}({\sqrt {2}}gk)}{k} \, . \label {sok}
\end{equation}
On the coefficients $S^{(0)}_{p}(g)$ the BES equation implies the
linear system,
\begin{eqnarray}
S^{(0)}_{2p}(g)&=&-4p\sum
_{m=1}^{\infty}Z_{2p,2m}(g)S^{(0)}_{2m}(g)+
4p\sum _{m=1}^{\infty}Z_{2p,2m-1}(g)S^{(0)}_{2m-1}(g)  \, , \nonumber \\
\label {S0eq} \\
S^{(0)}_{2p-1}(g)&=&2{\sqrt {2}}g\, \delta _{p,1}-2(2p-1)\sum
_{m=1}^{\infty}Z_{2p-1,2m}(g) S^{(0)}_{2m}(g)-2(2p-1)\sum
_{m=1}^{\infty}Z_{2p-1,2m-1}(g)S^{(0)}_{2m-1}(g) \nonumber \, ,
\end{eqnarray}
where we introduced the notation
\begin{equation}
\label{Zetaint}
Z_{n,m}(g)=\int _{0}^{+\infty} \frac {dh}{h} \frac {J_n({\sqrt
{2}}gh)J_m({\sqrt {2}}gh)}{e^h-1} \, .
\end{equation}
The cusp anomalous dimension can be extracted from the relations
\begin{equation}
\lim _{k\rightarrow 0^+}S^{(0)}(k)=\frac {1}{2}f(g) \, , \quad
f(g)={\sqrt {2}}g S^{(0)}_{1}(g) \, ,
\end{equation}
and its strong coupling behaviour was completely disentangled in
\cite {BKK}.

\subsection{On the first generalized scaling function}

The strong coupling limit of the first generalised scaling function was studied in \cite {FGR}.
Here we briefly recall the main results. We define the even function
\begin{equation}
S^{(1)}(k)= \frac {2\sinh \frac {|k|}{2}}{2\pi  |k|}\hat\sigma
_H^{(1)}(k) \label {Sdef} \, ,
\end{equation}
and introduce the two functions
\begin{equation}
a_r(g)=\int _{-\infty}^{+\infty}\frac {dh}{h} J_{r}({\sqrt
{2}}gh)\frac {1}{1+e^{\frac {|h|}{2}}} \, , \quad \bar a_r(g)=\int
_{-\infty}^{+\infty}\frac {dh}{|h|} J_{r}({\sqrt {2}}gh)\frac
{1}{1+e^{\frac {|h|}{2}}} \, .
\end{equation}
Expanding, for $k \geq 0$, in a series involving Bessel functions,
\begin{equation}
S^{(1)}(k)=\sum _{p=1}^{\infty}S_{2p}^{(1)}(g) \frac {J_{2p}({\sqrt
{2}}gk)}{k}+\sum _{p=1}^{\infty}S_{2p-1}^{(1)}(g)\frac
{J_{2p-1}({\sqrt {2}}gk)}{k} \, , \label {smexp}
\end{equation}
the coefficients $S_r^{(1)}(g)$ satisfy the linear system,
\begin{eqnarray}
S_{2p}^{(1)}(g)&=&2+2p \left ( -\bar a_{2p}(g)-2\sum _{m=1}^{\infty}Z_{2p,2m}(g)S_{2m}^{(1)}(g)+2\sum _{m=1}^{\infty}Z_{2p,2m-1}(g)S_{2m-1}^{(1)}(g) \right )\, ,  \nonumber \\
\label {Seq2} \\
\frac {S_{2p-1}^{(1)}(g)}{2p-1}&=&-a_{2p-1}(g)-2\sum
_{m=1}^{\infty}Z_{2p-1,2m}(g)S_{2m}^{(1)}(g)-2\sum
_{m=1}^{\infty}Z_{2p-1,2m-1}(g)S_{2m-1}^{(1)}(g) \nonumber \, .
\end{eqnarray}
In paper \cite {FGR} we found the following asymptotic strong
coupling solution to the system (\ref {Seq2}):
\begin{eqnarray}
S_{2m-1}^{(1)}(g)&\doteq&(2m-1)\sum _{n^{\prime}=1}^m (-1)^{n^{\prime}}
\frac {\Gamma (m+n^{\prime}-1)} {\Gamma (m-n^{\prime}+1)} \frac
{b_{2n^{\prime}-1}}{g^{2n^{\prime}-1}} \, , \nonumber \\
&& \label {sm} \\
S_{2m}^{(1)}(g)&\doteq&- 2m\sum _{n^{\prime}=1}^m (-1)^{n^{\prime}} \frac
{\Gamma (m+n^{\prime})} {\Gamma (m-n^{\prime}+1)} \frac
{b_{2n^{\prime}}}{g^{2n^{\prime}}} \, , \nonumber
\end{eqnarray}
where the coefficients
\begin{eqnarray}
b_{2n^{\prime}}&=&2^{-n^{\prime}}(-1)^{n^{\prime}} \sum
_{k=0}^{n^{\prime}} \frac
{E_{2k}2^{2k}}{(2k)!(2n^{\prime}-2k)!} \, , \nonumber \\
&& \label {aeul} \\
b_{2n^{\prime}-1}&=&2^{-n^{\prime}+\frac {1}{2}}(-1)^{n^{\prime}-1}
\sum _{k=0}^{n^{\prime}-1} \frac
{E_{2k}2^{2k}}{(2k)!(2n^{\prime}-2k-1)!} \, , \nonumber
\end{eqnarray}
with $E_{2k}$ the Euler's numbers, sum up to the generating function
\begin{equation}
b(t)=\sum _{n^{\prime}=0}^{\infty} b_{n^{\prime}} t^{n^{\prime}}
=\frac {1}{\cos \frac {t}{\sqrt {2}}-\sin \frac {t}{\sqrt {2}}} \, .
\end{equation}
In addition, the behaviour
\begin{eqnarray}
f_1(g)&=&2\lim _{k\rightarrow 0^+}S^{(1)}(k)={\sqrt {2}}g
S_1^{(1)}(g)=-1+m(g)+ O \left (e^{-3\frac
{\pi}{\sqrt {2}}g} \right ) \, ,  \nonumber \\
m(g)&=&\frac {2^{\frac {5}{8}}\pi
}{\Gamma \left ( \frac {5}{4} \right )} g^{\frac {1}{4}}e^{-\frac
{\pi g}{\sqrt {2}}}\left [1+\sum _ {n=1}^{\infty} \frac {k_n}{g^n}\right ] \, , \label {mgap}
 \end{eqnarray}
where $m(g)$ is the mass gap of the $O(6)$ nonlinear sigma model, expressed in terms of parameters of the underlying $\text{AdS}_5\times\text{S}^5$ sigma model, was shown (for the leading exponential)
in \cite{FGR}, after numerically solving the system (\ref {Seq2}). Later on, the embedding into the $O(6)$ NLSM has been proven analytically in \cite{BK}. In this perspective, the first massive excitations of the string theory give a natural cut-off which determines univocally the coefficients of the series, {\it i.e.}  $k_1,\, k_2, \dots $. These can be, in principle, computed analytically by using results of \cite{BK} and of the present paper (cf. (\ref{BKeq})); the first two, $k_1$, $k_2$, will be given a numerical estimate in subsection 5.2.

\subsection{On the second and higher generalised scaling functions}

We now give a general scheme for tackling the problem of computing
the $n$-th generalised scaling function $f_n(g)$ for $n\geq 2$ at
arbitrary value of the coupling constant.

We start from (\ref {sigmaeq2}) and define the function $S(k)$:
\begin{equation}
\ln s \ S(k)=\frac {2\sinh \frac {|k|}{2}}{2\pi |k|}\left [\hat
\sigma _H (k)+ \hat \sigma _0^s (k) \right ] +\frac {e^{-\frac
{|k|}{2}}}{\pi |k|} \int  _{-\infty }^{+\infty} \frac {dp}{2\pi}
\left [\hat \sigma _0^s(p)+\hat \sigma _H(p) \right ] \frac {\sin
(k-p)c} {k-p} \, , \label{Sndefi}
\end{equation}
which, differently from the cases $n=0$, $n=1$, depends on the all
loops density (\ref {allloopsdens}) $\hat \sigma (k)=\hat \sigma _H
(k)+ \hat \sigma _0^s (k)$. Since we are in the limit (\ref {jlimit}), we
naturally have
\begin{equation}
S(k)=s^{(0)}(k)+s^{(1)}(k)j+\sum _{n=2}^{\infty} S^{(n)}(k)j^n \, .
\label{defs}
\end{equation}
We focus on $S^{(n)}(k)$, with $n\geq 2$. For such
functions the following equation holds:
\begin{eqnarray}
&&S^{(n)}(k)= \label {sngeneq} \frac {1}{\pi |k| } \int _{-\infty
}^{+\infty} \frac {dh}{|h|} \Bigl [ \sum _{r=1}^{\infty} r
(-1)^{r+1}J_r({\sqrt {2}}gk) J_r({\sqrt {2}}gh)\frac {1-{\mbox
{sgn}}(kh)}{2}
e^{-\frac {|h|}{2}} + \nonumber \\
&+&{\mbox {sgn}} (h) \sum _{r=2}^{\infty}\sum _{\nu =0}^{\infty}
c_{r,r+1+2\nu}(g)(-1)^{r+\nu}e^{-\frac {|h|}{2}} \Bigl (
J_{r-1}({\sqrt {2}}gk) J_{r+2\nu}({\sqrt {2}}gh)- \label {snequ} \\
&-& J_{r-1}({\sqrt {2}}gh) J_{r+2\nu}({\sqrt {2}}gk)\Bigr ) \Bigr ]
\Bigl [\frac {\pi |h|}{\sinh \frac {|h|}{2}}S^{(n)}(h) -\frac
{e^{\frac {|h|}{2}}}{\sinh \frac {|h|}{2}}\int  _{-\infty
}^{+\infty} \frac {dp}{2\pi} \left . \hat \sigma (p)  \frac {\sin
(h-p)c} {h-p}  \right |_{j^n} \Bigr] \, , \nonumber
\end{eqnarray}
where the symbol $|_{j^n}$ means that we have to extract only the
coefficient of $j^n$ in the limit (\ref {jlimit}), after having
removed the overall factor $\ln s$.

Again, if we restrict the domain to $k \geq 0$ we can expand in
series of Bessel functions,
\begin{equation}
S^{(n)}(k)=\sum _{p=1}^{\infty}S^{(n)}_{2p}(g)\frac {J_{2p}({\sqrt
{2}}gk)}{k}+\sum _{p=1}^{\infty}S^{(n)}_{2p-1}(g)\frac
{J_{2p-1}({\sqrt {2}}gk)}{k} \, , \label {snbess}
\end{equation}
in such a way that the $n$-th generalised scaling function is
expressed as (\ref {egs2},\ref {Sndefi}):
\begin{equation}
f_n(g)={\sqrt {2}}g S^{(n)}_{1}(g) \, .
\end{equation}
After some computations (for details, see Appendix A of \cite {FGR2}),
we find the following system of equations
for the coefficients of $S^{(n)}(k)$, with $n\geq 2$,
\begin{eqnarray}
S^{(n)}_{2p}(g)&=&A_{2p}^{(n)}(g)-4p\sum _{m=1}^{\infty}Z_{2p,2m}(g)S^{(n)}_{2m}(g)+ 4p\sum _{m=1}^{\infty}Z_{2p,2m-1}(g)S^{(n)}_{2m-1}(g) \, , \nonumber \\
S^{(n)}_{2p-1}(g)&=&A_{2p-1}^{(n)}(g)-2(2p-1)\sum
_{m=1}^{\infty}Z_{2p-1,2m}(g)
S^{(n)}_{2m}(g)-\label {Seqn} \\
&-& 2(2p-1)\sum _{m=1}^{\infty}Z_{2p-1,2m-1}(g)S^{(n)}_{2m-1}(g)
\nonumber \, ,
\end{eqnarray}
where the 'forcing terms' $A_r^{(n)}(g)$ are given by:
\begin{equation}
A_r^{(n)}(g)=r \int _{0}^{+\infty}\frac {dh}{2\pi h} \,  \frac
{J_{r}({\sqrt {2}}gh)}{\sinh \frac {h}{2}} \, \left. \int
_{-\infty}^{+\infty} \frac {dp}{2\pi} 2 \frac {\sin (h-p)c }{h-p}
[\hat \sigma _0^s(p)+ \hat \sigma _H (p) ] \right |_{j^n} \, .
\label {forcrn}
\end{equation}
These systems have all the same kernel, which coincides with the BES
one, and differ only for their forcing terms. The inforcing of the
normalization conditions in (\ref {forcrn}) will show how the $n$-th
forcing term depends on the solutions of the $m$-th system, with
$m\leq n-3$, allowing, therefore, their iterative solution. This
will be the subject of next section, where we are going to
systematically tackle the problem of finding $A_r^{(n)}(g)$ for all
values of $n$, up to the desired order.

\medskip

As an example we now show that $\sigma _{H}^{(2)}(u)=0$, so that we obviously have
$f_2(g)=0$. Let us consider the r.h.s. of (\ref {sigmaeq}). The
first term is clearly proportional to $j\ln s$, so it does not
appear in the equation for $\sigma _{H}^{(2)}(u)$. The second and
the fifth term both have the form, with two different functions
$f(v)$,
\begin{equation}
\int  _{-\infty}^{+\infty} dv f(v)\sigma (v) \chi _{d} (v)= \int
_{-\infty}^{+\infty} \frac {dk}{2\pi} \hat f(k) \int
_{-\infty}^{+\infty} \frac {dp}{2\pi} \hat \sigma (p)
 2 \frac {\sin (k-p)d}{k-p} \, , \label {forc1}
\end{equation}
where $\sigma , d$ stand for $\sigma _0^s+\sigma _H, c$, respectively, if we
consider the second term, whilst for $\sigma _0^s, c_0$, respectively,  if we consider
the fifth term. Using the normalization condition
\begin{equation}
2 \int  _{-\infty }^{\infty} \frac {dp}{2\pi} \hat \sigma (p) \frac
{\sin pd}{p} =-2\pi (L-2) \, ,
\end{equation}
one can show that
\begin{equation}
2 \int  _{-\infty}^{+\infty} \frac {dp}{2\pi} \hat \sigma (p)
 \frac {\sin (k-p)d}{k-p}=[-2\pi j + O(d^3)]\ln s  \, .
\end{equation}
Since $d$ starts from order $j$ in its expansion, the second and the
fifth term in the r.h.s. of (\ref {sigmaeq}) lack of the order
$j^2\ln s $ terms in their expansion. The same reasoning, applied to
the second term in the rhs of (\ref {hatsigma0}) - the one
containing the integral - implies that also this term lacks of the
order $j^2\ln s$. Therefore, the third term in the r.h.s. of (\ref
{sigmaeq}) is missing the quadratic order as well. It follows that
the equation for $\sigma _{H}^{(2)}(u)$ is
\begin{eqnarray}
\sigma _{H}^{(2)}(u)&=& \int _{-\infty}^{+\infty} \frac {dv}{\pi}
\frac {1}{1+(u-v)^2}
\sigma _{H}^{(2)}(v) \\
&-& \frac {i}{\pi} \int _{-\infty}^{+\infty} dv \frac {d}{du} \left
[ \ln \left ( \frac {1-\frac {g^2}{2x^+(u)x^-(v)} }{1- \frac
{g^2}{2x^-(u)x^+(v)}} \right )+i \theta (u,v) \right ] \sigma
_{H}^{(2)}(v) \, , \nonumber
\end{eqnarray}
whose solution is, of course, $\sigma _{H}^{(2)}(u)=0$. Therefore
$f_2(g)=0$, as as already presented in the Bethe Ansatz \cite{FRS} and string (penultimate reference in \cite{CK}) literature.

\section{Systematorics}
\setcounter{equation}{0}

The main obstacle to obtain a fully explicit expression for the
infinite linear system at a generic order $n$ is the double
expansion contained in the term $\sin((h-p) c(j))$ of equation
(\ref{forcrn}). A similar structure is also present in the
normalization conditions (\ref{normale}).

In order to overcome this technical problem, it is worth to remember
a standard result of combinatorics known as the Fa\`a di Bruno's
formula \cite{faa}. Let $f(x)$ and $g(x)$ be a pair of functions
admitting (at least formally) a power expansion of this kind
\begin{eqnarray}
f(x) = \sum_{n=1}^{\infty} \frac{f_n}{n!} \, x^n, \ \ \ \ \ \ g(x) =
\sum_{n=1}^{\infty} \frac{g_n}{n!} \, x^n \, ,
\end{eqnarray}
then the composition $g(f(x))$ admits the following power expansion
\begin{eqnarray}
g(f(x)) =h(x)= \sum_{n=1}^{\infty} \frac{h_n}{n!} \, x^n \, ,
\end{eqnarray}
where the coefficients $h_n$ have the following form
\begin{eqnarray}
h_n = \sum_{k=1}^n g_k \, B_{n,k}(f_1, \dots, f_{n-k+1}) \, .
\end{eqnarray}
$B_{n,k} (\underline f)$ is the Bell polynomial defined as
\begin{eqnarray}
B_{n,k}(f_1, \dots, f_{n-k+1}) = n! \, \sum_{ \{j_1,\dots,j_{n-k+1}
\} } \prod_{m=1}^{n-k+1} \frac{(f_m)^{j_m}}{ j_m! \, (m!)^{j_m}}
\end{eqnarray}
and the sum runs over all the non negative $j$'s satisfying the
conditions
\begin{eqnarray}
\sum_{m=1}^{n-k+1} j_m  = k, \ \ \ \ \ \sum_{m=1}^{n-k+1} m \,j_m =
n. \nonumber
\end{eqnarray}
The previous equation will be our main tool in the remaining part of
this section. It is straightforward to apply the previous formula to
the present case, $\sin(p \, c(j))$, being
\begin{eqnarray}
\sin x = \sum_{n=1}^\infty \frac{{\xi}_n}{n!} x^n, \ \ \ \ {\xi}_n=
\frac12 \, i^{n+1} ((-1)^{n}-1)
\end{eqnarray}
and
\begin{eqnarray}
c(j)  = \sum_{n=1}^\infty c^{(n)} j^n.
\end{eqnarray}
We end up with (we divide by $p$ for future convenience)
\begin{eqnarray}
\frac{\sin(p \, c(j))}{p} &=& \sum_{n=1}^\infty \frac{\sum
_{k=1}^n{\xi}_k}{p \,n!} \, B_{n,k}(p \, c^{(1)}, \dots, p
\,(n-k+1)! \, c^{(n-k+1)})\; j^n =
\nonumber \\
&=&\sum_{n=1}^\infty \Lambda_n(p) \, j^n.
\end{eqnarray}
Let us now use this result in order to write in a more convenient
way both the normalization conditions (\ref{normale}) and the forcing
term (\ref{forcrn}). We begin with the analysis of $\Lambda_n(p)$ in
order to put it in a more suitable form. Some elementary
manipulations give the formula
\begin{eqnarray}
\Lambda_n(p) = \sum_{k=1}^n {\xi}_k \, \beta_{n,k} (c^{(1)}, \dots ,
c^{(n-k+1)};p) \, , \label {lambdan}
\end{eqnarray}
with
\begin{eqnarray}
&& \beta_{n,k} (c^{(1)}, \dots , c^{(n-k+1)};p)= \, \sum_{ \{j_1,\dots,j_{n-k+1} \} } (p^{k-1})\, \prod_{m=1}^{n-k+1} \frac{(c^{(m)})^{j_m}}{ j_m! \, } \, , \nonumber \\
&&  \sum_{m=1}^{n-k+1} j_m = k \, , \quad  \sum_{m=1}^{n-k+1}mj_m =
n \, .
\end{eqnarray}
The forcing term and the normalization conditions have a common
structure
\begin{eqnarray}
\frac{\sin(p_1 \, c(j))}{p_1} \hat{\sigma}(p_2) &=& \left( \sum_{n=1}^\infty \Lambda_n(p_1) \, j^n \right)\left( \sum_{n=0}^\infty  \hat{\sigma}^{(n)}(p_2) \, j^n \right)\ln s = \nonumber \\
&=& \sum_{n=1}^\infty \Gamma_n(p_1,p_2) \, j^n \ln s \, ,
\end{eqnarray}
where
\begin{eqnarray}
 \Gamma_n(p_1,p_2) =
\sum_{k=1}^n \Lambda_k(p_1) \, \hat{\sigma}^{(n-k)}(p_2).
\end{eqnarray}
We can finally write the coefficient in the expansion in powers of
$j$ of the integral over the momentum which appears in the forcing
term as
\begin{eqnarray}
2  \int_{-\infty}^{+\infty} \frac{dp}{2 \pi} \, \Gamma_n(h-p,p)  \ln
s \, ,
\end{eqnarray}
along with the normalization condition
\begin{eqnarray}
\int_{-\infty}^{+\infty} \frac{dp}{2 \pi} \, \Gamma_n(p,p) = - \pi
\, \delta_{n,1} . \label{normy}
\end{eqnarray}
Our next step will be to enforce the normalization condition in the
forcing term in order to gain a simplification of its structure. We
notice that $ \Lambda _n(p)$ has a momentum independent term
corresponding to the term $k=1$ in the sum (\ref {lambdan}) and
hence $ \Gamma_n(p_1,p_2)$ admits the following decomposition
\begin{eqnarray}
 \Gamma_n(p_1,p_2) =  \Gamma^{(0)}_n(p_2) + \tilde{\Gamma}_n(p_1,p_2),
 \ \ \ \  \Gamma^{(0)}_n(p_2) = \sum_{k=1}^n \hat{\sigma}^{(n-k)}(p_2) \, c^{(k)}.
\end{eqnarray}
The normalization condition then becomes
\begin{eqnarray}
-\int_{-\infty}^{+\infty} \frac{dp}{2 \pi} \, \Gamma^{(0)}_n(p) =
\int_{-\infty}^{+\infty} \frac{dp}{2 \pi} \, \tilde{\Gamma}_n(p,p)
+\pi \, \delta_{n,1} \, ,
\end{eqnarray}
which allows to subtract the $\Gamma^{(0)}_n(p)$ contribution in the
forcing term:
\begin{eqnarray}
&& \nonumber 2 \int_{-\infty}^{+\infty}\frac{dp}{2 \pi} \, \Gamma_n(h-p,p) = \\
&& = -2 \pi \delta_{n,1} + 2 \int_{-\infty}^{+\infty}\frac{dp}{2
\pi} \, \left[ \Gamma_n(h-p,p) -\Gamma_n(p,p) \right]  \\ \nonumber
&& = -2 \pi \delta_{n,1} + 2 \int_{-\infty}^{+\infty}\frac{dp}{2
\pi} \, \left[ \Delta_n(h-p,p) \right] \, .
\end{eqnarray}
For instance, this subtraction is responsible for $A_r^{(2)}(g) = 0$ and hence for what has been noticed at end of sub-section 3.3, {\it i.e.} $f_2(g)=0$.

The final step is to make this subtraction explicit for any $n$. With this aim in mind, we need to consider the even parity of the Fourier transforms of the densities $\hat{\sigma}^n(p)$ and we define the $s$-derivative of the $n$-th density in $u=0$ as
\be
\sigma^{(n);(s)}\equiv \frac{d^s
\sigma^{(n)}(u)}{du^s}\Big|_{u=0} \, .
\label{defsigma}
\ee
Thus, the following relation allows us to perform the integral over $p$,
\begin{eqnarray}
i^{-s}  \sigma^{(n-k);(s)}&=& \int_{-\infty}^{+\infty}
\frac{dp}{2 \pi} \, p^s \, \hat{\sigma}^{(n-k)}(p) = \nonumber \\
&=& 2 d_s \int_{0}^{+\infty}
\frac{dp}{2 \pi} \, p^s \, \hat{\sigma}^{(n-k)}(p) \, ,  \quad
d_s =\frac12 (1+(-1)^s) \, , \label{FTderiv}
\end{eqnarray}
which is different from zero only for $s$ even, due to the parity
property of $\hat{\sigma}^{(n-k)}(p)$.

It is then possible to rewrite $\Delta_n(h-p,p) $ as follows
\begin{eqnarray}
\Delta_n(h-p,p) = \sum_{k=1}^n \hat{\sigma}^{(n-k)}(p) (
\Lambda_k(h-p) - \Lambda_k(p) ) \, ,
\end{eqnarray}
where
\begin{eqnarray}
&& \Lambda_k(h-p) - \Lambda_k(p) =  \sum_{l=1}^k {\xi}_l \, \sum_{
\{j_1,\dots,j_{k-l+1} \} } \prod_{m=1}^{k-l+1}
\frac{(c^{(m)})^{j_m}}{ j_m! } \left( (h-p)^{l-1} - p^{l-1}
\right) \nonumber \\
&& = \sum_{l=1}^k {\xi}_l \, \sum_{ \{j_1,\dots,j_{k-l+1} \} }  \prod_{m=1}^{k-l+1} \frac{(c^{(m)})^{j_m}}{ j_m! } \left( \sum_{s=0}^{l-1} \left(\begin{array}{c}l-1 \\s\end{array}\right)h^{l-1-s} (-1)^s p^s - p^{l-1} \right) \nonumber \\
&& = \sum_{l=3}^k {\xi}_l \, \sum_{ \{j_1,\dots,j_{k-l+1} \} }  \prod_{m=1}^{k-l+1} \frac{(c^{(m)})^{j_m}}{ j_m! } \left( \sum_{s=0}^{l-2} \left(\begin{array}{c}l-1 \\s\end{array}\right)h^{l-1-s} (-1)^s p^s + (-p)^{l-1}- p^{l-1} \right) \nonumber \\
&& = \sum_{l=3}^k {\xi}_l \,  \left( \sum_{s=0}^{l-2} \left(\begin{array}{c}l-1 \\s\end{array}\right)h^{l-1-s} (-p)^s  \right) \,\sum_{ \{j_1,\dots,j_{k-l+1} \} }  \prod_{m=1}^{k-l+1} \frac{(c^{(m)})^{j_m}}{ j_m! } \, , \\
&& \sum_{m=1}^{k-l+1} j_m  = l, \ \ \ \ \ \sum_{m=1}^{k-l+1} m \,j_m
= k \, . \nonumber
\end{eqnarray}
The last step comes from the fact that it is always $(-p)^{l-1}-
p^{l-1} =0 $, because ${\xi}_l$ is non-vanishing only for odd $l$.
One can also notice that the subtraction and the fact that
${\xi}_2=0$ allow the sum over $l$ to begin from $l=3$.

The previous result, together with eq. (\ref{FTderiv}), allows to
write down, for $n\geq 2$,
\begin{eqnarray}
&& 2 \int_{-\infty}^{+\infty}\frac{dp}{2 \pi} \, \Gamma_n(h-p,p) =2
\int_{-\infty}^{+\infty} \frac{dp}{2 \pi} \, \left[ \Gamma_n(h-p,p)
-\Gamma_n(p,p) \right] \label{jexp}   \\ \nonumber && = 2
\int_{-\infty}^{+\infty} \frac{dp}{2 \pi} \, \left[ \Delta_n(h-p,p)
\right] = \\ \nonumber && =  2  \sum_{k=1}^n  \sum_{l=3}^k  {\xi}_l
\, \left( \sum_{s=0}^{l-2} \left(\begin{array}{c}l-1
\\s\end{array}\right) d_s \, h^{l-1-s} (-i)^{-s} \sigma^{(n-k);(s)}
\right) \, \sum_{ \{j_1,\dots,j_{k-l+1} \} }  \prod_{m=1}^{k-l+1}
\frac{(c^{(m)})^{j_m}}{ j_m! } \, ,
\\
&&  \sum_{m=1}^{k-l+1} j_m  = l, \ \ \ \ \ \sum_{m=1}^{k-l+1} m
\,j_m = k \, \nonumber \, ,
\end{eqnarray}
which is nothing but the explicit $n$-th term of the $j$ expansion
of the integral over $p$ which appears in the forcing term. Then, if
we pose
\begin{eqnarray}
\mathbb{I}_{r}^{l,s} = r \int_0^{+\infty} \frac{dh}{2 \pi h}
\frac{J_r(\sqrt2 g h)}{\sinh \frac{ h}{2}} h^{l-1-s} \, , \nonumber
\end{eqnarray}
we can explicitly write down the generic expression for the forcing
term $A_r^{(n)}(g)$, $n\geq 2$, entering the system (\ref {Seqn}) for
the $n$-th term of the $j$ expansion of the function $S(k)$:
\begin{eqnarray}
&& A_r^{(n)}(g) =  \label {forterm} \\
&& = 2  \sum_{k=1}^n  \sum_{l=3}^k  {\xi}_l \,  \left( \sum_{s=0}^{l-2} \left(\begin{array}{c}l-1 \\s\end{array}\right) d_s \, (-i)^{-s} \sigma^{(n-k);(s)} \, \mathbb{I}_{r}^{l,s} \right) \, \sum_{ \{j_1,\dots,j_{k-l+1} \} }  \prod_{m=1}^{k-l+1} \frac{(c^{(m)})^{j_m}}{ j_m! }, \nonumber \\
&&  \sum_{m=1}^{k-l+1} j_m  = l, \ \ \ \ \ \sum_{m=1}^{k-l+1} m
\,j_m = k \, .\nonumber
\end{eqnarray}
Because of the particular form of the forcing terms, it is
convenient to write the solution of (\ref {Seqn}) as
\begin{eqnarray}
&& S_r^{(n)}(g) =  \label {forterm22} \\
&& = 2  \sum_{k=1}^n  \sum_{l=3}^k  {\xi}_l \,  \left( \sum_{s=0}^{l-2} \left(\begin{array}{c}l-1 \\s\end{array}\right) d_s \, (-i)^{-s} \sigma^{(n-k);(s)} \, \tilde S_{r}^{\left (\frac {l-s-1}{2}\right ) } (g)\right) \, \sum_{ \{j_1,\dots,j_{k-l+1} \} }  \prod_{m=1}^{k-l+1} \frac{(c^{(m)})^{j_m}}{ j_m! }, \nonumber \\
&&  \sum_{m=1}^{k-l+1} j_m  = l, \ \ \ \ \ \sum_{m=1}^{k-l+1} m
\,j_m = k \, ,\nonumber
\end{eqnarray}
where the ``reduced'' coefficients $\tilde S_r^{(k)}$ satisfy the
equations
\begin{eqnarray}
\tilde S^{(k)}_{2p}(g)&=&{\mathbb I}_{2p}^{(k)}(g)-4p\sum _{m=1}^{\infty}Z_{2p,2m}(g)\tilde S^{(k)}_{2m}(g)+ 4p\sum _{m=1}^{\infty}Z_{2p,2m-1}(g)\tilde S^{(k)}_{2m-1}(g) \, ,  \nonumber \\
\tilde S^{(k)}_{2p-1}(g)&=&{\mathbb I}_{2p-1}^{(k)}(g)- 2(2p-1)\sum
_{m=1}^{\infty}Z_{2p-1,2m}(g)
\tilde S^{(k)}_{2m}(g)-\label {redSeqn2} \\
&-& 2(2p-1)\sum _{m=1}^{\infty}Z_{2p-1,2m-1}(g)\tilde
S^{(k)}_{2m-1}(g) \nonumber \, ,
\end{eqnarray}
with the reduced forcing terms,
\begin{equation}
\label{intforterm}
{\mathbb I}_r^{(k)}= r \int _{0}^{+\infty} \frac {dh}{2\pi} h^{2k-1}
\frac {J_r ({\sqrt {2}}gh)}{\sinh \frac {h}{2}} \, ,
\end{equation}
which are 'known' functions, {\it i.e.} they do not depend on the
quantities $\sigma ^{(n');(s)}$. We notice that inside the structure
of the forcing term $A_r^{(n)}(g)$ (\ref {forterm}) we find the
constants $c^{(m)}$, with $m\leq n-2$ and the densities of the Bethe
roots at $u=0$ (together with their derivatives) $\sigma
^{(n');(s)}$, with $n'\leq n-3$. In addition, the constants
$c^{(m)}$ can be related to $\sigma ^{(n');(s)}$, with $n'\leq m-1$,
by means of the normalization condition (\ref{normy}), thus leaving
the forcing term $A_r^{(n)}(g)$ as dependent on $\sigma
^{(n');(s)}$, with $n'\leq n-3$. Let us now find the relation
between $c^{(m)}$ and $\sigma ^{(n');(s)}$. We first of
all notice that, for $n=1$, we have
\begin{eqnarray}
c^{(1)} = - \frac{\pi}{\sigma^{(0);(0)}} \label{inicond}
\end{eqnarray}
and that, for $m>1$, the normalization condition takes the form
\begin{eqnarray}
&&\sum_{k=1}^m  \sum_{l=1}^k  {\xi}_l \, (i)^{-l+1} \sigma^{(m-k);(l-1)}  \, \sum_{ \{j_1,\dots,j_{k-l+1} \} }  \prod_{m'=1}^{k-l+1} \frac{(c^{(m')})^{j_{m'}}}{ j_{m'}! }=0 \, , \\
&& \sum_{m'=1}^{k-l+1} j_{m'}  = l, \ \ \ \ \ \sum_{m'=1}^{k-l+1} m'
\,j_{m'} = k \, . \nonumber
\end{eqnarray}
After a brief inspection of the latter it is possible to realize
that, at order $m$, the only term which contains $c^{(m)}$ can be
singled out by taking $k=m$, $l=1$, and that the remaining terms in
the sums only contain $c^{(k)}$ with $k<m$.

As a consequence, we can write a recursion relation
\begin{eqnarray}
c^{(m)} & = &-  \sum_{k=1}^{m-1}
\frac{\sigma^{(m-k);(0)}}{\sigma^{(0);(0)}}  c^{(k)} - \nonumber
\\
&&- \sum_{k=1}^m  \sum_{l=2}^k  {\xi}_l \, (i)^{-l+1} \frac{\sigma^{(m-k);(l-1)}}{\sigma^{(0);(0)}} \sum_{ \{j_1,\dots,j_{k-l+1} \} }  \prod_{m'=1}^{k-l+1} \frac{(c^{(m')})^{j_{m'}}}{ j_{m'}! }, \ \ \ m>1 \label{recurcoeff} \, , \\
&& \sum_{m'=1}^{k-l+1} j_{m'}  = l, \ \ \ \ \ \sum_{m'=1}^{k-l+1} m'
\,j_{m'} = k \, . \nonumber
\end{eqnarray}
which, together with the initial condition (\ref{inicond}), allows
to express all the $c^{(m)}$ recursively, in terms of $\sigma
^{(n');(s)}$, with $n'\leq m-1$. Therefore, we conclude that the
forcing term $A_r^{(n)}(g)$ (\ref {forterm}) and the solution
$S_r^{(n)}(g)$ (\ref {forterm22}) actually depend only on $\sigma
^{(n');(s)}$, with $n'\leq n-3$, {\it i.e.} on the solutions of previous
systems. Consequently, at least in principle, the solution for the
$S_r^{(n)}(g)$ may be found by recursive methods.

\medskip

To summarise, the principal result of this section is formula (\ref
{forterm22}): the evaluation of the $n$-th generalised scaling
function $f_n(g)={\sqrt {2}}g S^{(n)}_{1}(g)$, for $n\geq 2$, is
eventually reduced to the knowledge of $\tilde S^{(k)}_{1}(g)$ and
of the densities and their derivatives in zero, $\sigma ^{(n');(s)}$
(\ref {FTderiv}), with $n'\leq n-3$. In next subsection, we will
show that $\tilde{S}^{(k)}_1(g) $ (and $f_1(g)={\sqrt {2}}g
S^{(1)}_{1}(g)$) can be given an integral representation in terms of
the solution of the BES equation. However, this connection to the
BES equation (true, for obvious reasons, also for $\sigma
^{(0);(s)}$) is not true for the densities and their derivatives at
zero $\sigma ^{(n');(s)}$, $n'\geq 1$: in order to find them, one
needs more additional information, {\it i.e.} the full solution
$S^{(1)}_{r}(g)$, $\tilde S^{(k)}_{r}(g)$, for all $r$, to the
systems (\ref {Seq2}), (\ref {redSeqn2}). However, we again stress
that, due to iterative structure of (\ref {forterm22}), an explicit
solution for the $S_r^{(n)}(g)$ can be found by recursive
methods. This will be explicitly shown in the strong coupling limit
(section 5).

\subsection{Mapping the reduced systems to the BES equation}

As stated before, the main point of this subsection is to write down
an integral representation for the reduced coefficient
$\tilde{S}^{(k)}_1(g) $ and for $S^{(1)}_{1}(g)$, in terms of the
solution of the BES equation.

As a first step we rewrite the BES linear system (\ref{S0eq})
introducing the even/odd Neumann expansion\footnote{The use of
$\sigma _{\pm}^{(0)} (\sqrt{2} g t)$ is redundant with respect to
$S^{(0)}(k)$ (\ref {sok}). However, since in Appendix A we will use
results of \cite {BK}, we prefer to use here notations of that paper.}
\begin{eqnarray}
\sigma_+^{(0)} (\sqrt{2} g t)=  \sum_{p=1}^\infty \, S^{(0)}_{2p}(g)
\, J_{2p}(\sqrt{2} g t)  , \ \ \ \ \sigma_-^{(0)} (\sqrt{2} g t)=
\sum_{p=1}^\infty \, S^{(0)}_{2p-1}(g) \, J_{2p-1}(\sqrt{2} g t) \,
, \label {sgt}
\end{eqnarray}
with the coefficients $ S^{(0)}_{r}(g)$ given by
\begin{eqnarray}
S^{(0)}_{2p}(g) = 2 (2p) \, \int_0^{+\infty} \frac{dt}{t}
\sigma_+^{(0)} (t) J_{2p}(t)  , \ \ \ \ S^{(0)}_{2p-1}(g) = 2 (2p-1)
\, \int_0^{+\infty} \frac{dt}{t} \sigma_-^{(0)} (t) J_{2p-1}(t) \, .
\end{eqnarray}
Then, the BES linear system can be cast in the form \cite{BKK}
\begin{eqnarray}
&& \int_0^{+\infty} \frac{dt}{t} \left[ \frac{\sigma_+^{(0)} (\sqrt{2} g t)}{1-e^{-t}} -  \frac{\sigma_-^{(0)} (\sqrt{2} g t)}{e^{t}-1} \right] J_{2p}(\sqrt{2} g t) = 0 \, , \nonumber \\
\label{sysbes} \\
&&  \int_0^{+\infty} \frac{dt}{t} \left[ \frac{\sigma_-^{(0)}
(\sqrt{2} g t)}{1-e^{-t}} +  \frac{\sigma_+^{(0)} (\sqrt{2} g
t)}{e^{t}-1} \right] J_{2p-1}(\sqrt{2} g t) =  \sqrt{2} g \,
\delta_{1,p} \, . \nonumber
\end{eqnarray}
Since the kernel of the reduced system (\ref {redSeqn2}) is the same
as the BES one (\ref {S0eq}), it is possible to use the same
procedure introducing the functions
\begin{eqnarray}
\sigma_+^{(k)} (\sqrt{2} g t)=  \sum_{p=1}^{+\infty} \,
\tilde{S}^{(k)}_{2p} (g)\, J_{2p}(\sqrt{2} g t)  , \ \ \ \
\sigma_-^{(k)} (\sqrt{2} g t)= \sum_{p=1}^{+\infty} \,
\tilde{S}^{(k)}_{2p-1}(g)\, J_{2p-1}(\sqrt{2} g t)\, ,
\end{eqnarray}
together with
\begin{eqnarray}
\tilde{S}^{(k)}_{2p} (g)= 2 (2p) \, \int_0^{\infty} \frac{dt}{t}
\sigma_+^{(k)} (t) J_{2p}(t)  , \ \ \ \ \tilde{S}^{(k)}_{2p-1} (g)=
2 (2p-1) \, \int_0^{\infty} \frac{dt}{t} \sigma_-^{(k)} (t)
J_{2p-1}(t) \, .
\end{eqnarray}
And, from the system (\ref{redSeqn2}), we derive the following
equations for the functions $ \sigma_\pm^{(k)} (t)$:
\begin{eqnarray}
&& \int_0^{+\infty} \frac{dt}{t} \left[ \frac{\sigma_+^{(k)}
(\sqrt{2} g t)}{1-e^{-t}} -  \frac{\sigma_-^{(k)} (\sqrt{2} g
t)}{e^{t}-1} \right] J_{2p}(\sqrt{2} g t) = \frac{1}{4 \pi}
\int_0^{+\infty} dt \frac{t^{2k-1}}{\sinh t/2} J_{2p}(\sqrt{2} g t)
\, ,
\nonumber \\
\label{sysrid} \\
&&  \int_0^{+\infty} \frac{dt}{t} \left[ \frac{\sigma_-^{(k)}
(\sqrt{2} g t)}{1-e^{-t}} +  \frac{\sigma_+^{(k)} (\sqrt{2} g
t)}{e^{t}-1} \right] J_{2p-1}(\sqrt{2} g t) =  \frac{1}{4 \pi}
\int_0^{+\infty} dt \frac{t^{2k-1}}{\sinh t/2} J_{2p-1}(\sqrt{2} g
t) . \nonumber
\end{eqnarray}
The next step is to perform some manipulations on systems (\ref
{sysbes}, \ref {sysrid}), in order to exploit their similarities.
Concentrating first on (\ref {sysbes}), we multiply both sides of
the first equation by ${\tilde S}^{(k)}_{2p}(g)$, and both sides of
the second equation by ${\tilde S}^{(k)}_{2p-1}(g)$. Summing over
$p$ in both of them, we end up with
\begin{eqnarray}
&& \int_0^{+\infty} \frac{dt}{t} \left[ \frac{\sigma_+^{(0)} (\sqrt{2} g  t) \, \sigma_+^{(k)} (\sqrt{2} g t)}{1-e^{-t}} -  \frac{\sigma_-^{(0)} (\sqrt{2} g t) \, \sigma_+^{(k)} (\sqrt{2} g t)}{e^{t}-1} \right] = 0 \, , \nonumber \\
&&  \int_0^{+\infty} \frac{dt}{t} \left[  \frac{\sigma_-^{(0)}
(\sqrt{2} g t) \, \sigma_-^{(k)} (\sqrt{2} g t)}{1-e^{-t}} +
\frac{\sigma_+^{(0)} (\sqrt{2} g t) \, \sigma_-^{(k)} (\sqrt{2} g
t)}{e^{t}-1}  \right]  =  \sqrt{2} g \, \tilde{S}^{(k)}_{1} (g) \, ,
\nonumber
\end{eqnarray}
where we notice that the coefficient $\tilde{S}^{(k)}_{1}(g)$ is
explicitly singled out.

The same procedure can be repeated upon (\ref{sysrid}), by
multiplying the first equation by $S^{(0)}_{2p}(g)$, the second by
$S^{(0)}_{2p-1}(g)$ and finally summing over $p$. The result is as
follows:
\begin{eqnarray}
&& \int_0^{+\infty} \frac{dt}{t} \left[ \frac{\sigma_+^{(0)}
(\sqrt{2} g t) \, \sigma_+^{(k)} (\sqrt{2} g t)}{1-e^{-t}} -
\frac{\sigma_+^{(0)} (\sqrt{2} g t) \, \sigma_-^{(k)} (\sqrt{2} g
t)}{e^{t}-1} \right] = \frac{1}{4 \pi}
\int_0^{+\infty} dt \frac{t^{2k-1}}{\sinh t/2} \sigma^{(0)}_+(\sqrt{2} g t) \, ,  \nonumber \\
&&  \int_0^{+\infty} \frac{dt}{t} \left[  \frac{\sigma_-^{(0)}
(\sqrt{2} g t) \, \sigma_-^{(k)} (\sqrt{2} g t)}{1-e^{-t}} +
\frac{\sigma_-^{(0)} (\sqrt{2} g t) \, \sigma_+^{(k)} (\sqrt{2} g
t)}{e^{t}-1}  \right]  = \frac{1}{4 \pi} \int_0^{+\infty} dt
\frac{t^{2k-1}}{\sinh t/2} \sigma^{(0)}_-(\sqrt{2} g t)    \nonumber
\, .
\end{eqnarray}
A direct comparison with the previous equations allows to eventually
obtain the integral representation for $\tilde{S}^{(k)}_{1}(g)$,
\begin{eqnarray}
\sqrt{2} g \, \tilde{S}^{(k)}_{1}(g)  = - \frac{1}{4 \pi}
\int_0^{+\infty} dt \frac{t^{2k-1}}{\sinh t/2}
\left[\sigma^{(0)}_+(\sqrt{2} g t)  - \sigma^{(0)}_-(\sqrt{2} g t)
\right] \, . \label{intrid}
\end{eqnarray}
For what concerns the coefficient $S_1^{(1)}(g)$, relevant for the
computation of $f_1(g)$, the procedure is identical - one starts
from (\ref  {Seq2}) - but the result is slightly different. We have
\begin{eqnarray}
\sqrt{2} g \, {S}_{1}^{(1)} (g)   = - \int_0^{\infty} \frac {dt}{t}
\frac{1}{2\cosh t/4} \left[e^{-\frac {t}{4}}\sigma^{(0)}_-(\sqrt{2}
g t)  + e^{\frac {t}{4}} \sigma^{(0)}_+(\sqrt{2} g t) \right] \, .
\label{intone}
\end{eqnarray}
Equations (\ref {intrid}, \ref {intone}) are the representations of,
respectively, $\tilde{S}^{(k)}_{1}(g)$ and $S_1^{(1)}(g)$ in terms of
the BES quantities $\sigma^{(0)}_{\pm }(\sqrt{2} g t)$, defined in (\ref {sgt}).

Now, for completeness' sake and since we will be using it, we write
also the integral representation in terms of the solution of the BES
equation for the density and its derivatives in zero
$\sigma^{(0);(s)}$. This representation follows trivially from
the definition (\ref {S0def}):
\begin{eqnarray}
i^{-s}  \sigma^{(0);(s)}&=&-4\delta _{s,0} + i^{-s}  \sigma^{(0);(s)}_H \, , \nonumber \\
i^{-s}  \sigma^{(0);(s)}_H&=& d_{s}\int_0^{+\infty} dk \,
\frac{k^{s} }{\sinh k/2} \,
\sum_{p=1}^\infty S^{(0)}_p(g) \, J_p(\sqrt{2} g k) = \label{intdens} \\
&=& d_{s}\int_0^{+\infty} dk \, \frac{k^{s} }{\sinh k/2} \,
\left[\sigma^{(0)}_+(\sqrt{2} g k)  + \sigma^{(0)}_-(\sqrt{2} g k)
\right] \, . \nonumber
\end{eqnarray}
We have reached naturally one of the main points of our work and
some comments are in order. We have found that the generalised
scaling functions $f_n(g)$, $n\geq 2$, enjoy an expression (\ref
{forterm22}) in terms of the 'reduced' coefficients
$\tilde{S}^{(k)}_{1}(g)$ - related to the solution of the BES
equation - and of the densities and their derivatives at zero
$\sigma ^{(n');(s)}$, $n'\leq n-3$ - related (when $n' \geq 1$) also
to the other, 'higher', systems of equations (\ref {Seq2}), (\ref
{redSeqn2}). This result is in close analogy with formula (A.4) of \cite {BK}:
this is an integral expression in which the generalised scaling function is given
in terms of solutions of the BES equations and of the density of holes, which in the limit (\ref {jlimit})
is expressed in terms of the various $\sigma ^{(n');(s)}$. Our new contribution in the subject
is to have highlighted the recursive structure of general formul{\ae}
for $f_n(g)$: this allows analytic and numerical evaluations, which we will perform
explicitly in the strong coupling limit.

From the physical point of view the quantities
$\tilde{S}^{(k)}_{1}(g)$ and $\sigma ^{(n');(s)}$ can be reorganised in order to define different 'masses'
of the theory, denoted below by $m_n(g)$ (the precise definition of the $m_n(g)$ will be given on particular
examples at the end of subsection 5.1).
While computations of the various $\tilde{S}^{(k)}_{1}(g)$ and $\sigma ^{(n');(s)}$
at generic $g$ is technically challenging,
in next section we will show that all these independent quantities can be explicitly computed in
the strong coupling limit, by exploiting the recursive properties of the equations they satisfy.
Importantly, we will show that - confirming predictions contained in \cite {AM} - in the strong coupling limit they will be all expressed in terms of one quantity, the mass gap $m(g)$ (\ref {mgap}) of the $O(6)$ nonlinear sigma model, embedded in the ${\cal N}=4$ SYM theory, which is, in particular, the limiting value of all the $m_n(g)$.
In our approach it will be also possible to study their
corrections to the pure $O(6)$ limit: results on this issue are contained in subsection 5.2.

\section{Explicit results at strong coupling}
\setcounter{equation}{0}

In this section  we will solve the system (\ref {redSeqn2}) for $\tilde S_{r}^{(k)}(g)$ as the asymptotic series in inverse powers of $g$. Also, we will compute the leading strong coupling order of two
sets of quantities: on the one hand $\sigma ^{(0);(s)}$ and $\tilde
S_1^{(k)}(g)$, on the other hand $\sigma ^{(n');(s)}$, with $n'\geq
1$, and $c^{(m)}$ \footnote{In this second case we will also consider non-analytic corrections, which which are here of exponential nature. Since these corrections do not have an asymptotic expansion, they are called sometimes non-asymptotic.} .

The first set of quantities can be computed by relying on the
solution of the BES problem. Indeed, a simple but long calculation, whose
details are given in Appendix A, allows us to give the following
analytic estimates for large $g$:
\begin{eqnarray}
\sqrt{2} g \tilde{S}^{(k)}_1 (g)& = & \frac{(-1)^{k+1}}{4 \pi}
\left( \frac{\pi}{2} \right)^{2k} \, 2\, m(g)+ O\left (e^{-\frac{3\pi g
}{ \sqrt{2} }}\right ) \, ,
\nonumber \\
\sigma_H^{(0);(2k)} &=&  - \left( \frac{\pi}{2} \right)^{2k} \, \pi
\, m (g) + O\left (e^{-\frac{3\pi g}{ \sqrt{2} }}\right ) \, , \ \ \ \ \ \ k>0
\, ,
\label {cSm0} \\
\sigma_H^{(0);(0)} & = &   4 - \pi \, m (g) + O\left (e^{-\frac{3\pi g}{
\sqrt{2} }}\right )\, , \nonumber
\end{eqnarray}
where $m(g)$ (\ref {mgap}) turns out to be the $O(6)$ NLSM mass gap. Using the first of (\ref {intdens}) we can write, for the all loop density,
\begin{eqnarray}
\sqrt{2} g \tilde{S}^{(k)}_1 (g)& = & \frac{(-1)^{k+1}}{4 \pi}
\left( \frac{\pi}{2} \right)^{2k} \, 2\, m(g)+ O\left (e^{-\frac{3 \pi
g}{ \sqrt{2} }}\right ) \, ,
\nonumber \\
\label {cSm} \\
\sigma ^{(0);(2k)} &=&  - \left( \frac{\pi}{2} \right)^{2k} \, \pi
\, m (g) + O\left (e^{-\frac{3\pi g}{ \sqrt{2} }}\right ) \, , \nonumber
\end{eqnarray}
thus showing that in the strong coupling limit the quantities
$\sigma ^{(0);(s)}$ and $\tilde
S_1^{(k)}(g)$ (which, for generic $g$, are 'independent' quantities) flow into expressions depending on the single function $m(g)$.

For what concerns the $\sigma ^{(1);(s)}$, using the asymptotic
solution (\ref {sm}, \ref {aeul}) for $S_p^{(1)}(g)$,
we will explicitly write below in (\ref {dens1}) their strong coupling leading term.

Finally, in order to deal with the $\sigma ^{(n');(s)}$, with $n'\geq 2$, and the
$c^{(m)}$, we will first solve the system (\ref {redSeqn2}) for $\tilde S_{r}^{(k)}(g)$ as the asymptotic series in inverse powers of $g$.
Then, we will plug this expression into the recursive equation (\ref
{signs}) relating the quantities $\sigma
^{(n');(s)}$, with $n'\geq 2$, and the constants $c^{(m)}$. This
equation - together with (\ref{recurcoeff}) - will allow to find
recursively the strong coupling behaviour of both  $\sigma ^{(n');(s)}$, with $n'\geq 2$, and the
$c^{(m)}$.

Let us start from the reduced system (\ref {redSeqn2}) and let us look for a solution of it in the
form
\begin{equation}
\tilde S_{2m}^{(k)}(g)\doteq\sum _{n=k}^{\infty} \frac {\tilde
S_{2m}^{(k;2n)}}{g^{2n}} \, , \quad \tilde S_{2m-1}^{(k)}(g)\doteq \sum
_{n=k+1}^{\infty} \frac {\tilde S_{2m-1}^{(k;2n-1)}}{g^{2n-1}} \, ,
\end{equation}
with
\begin{eqnarray}
\tilde S_{2m}^{(k;2n)}&=&2m \frac {\Gamma (m+n)}{\Gamma (m-n+1)}(-1)^{1+n}\tilde b_{2n}^{(k)} \, , \quad n \geq k \, , \\
\tilde S_{2m-1}^{(k;2n-1)}&=&(2m-1) \frac {\Gamma (m+n-1)}{\Gamma
(m-n+1)}(-1)^{n}\tilde b_{2n-1}^{(k)} \, , \quad n \geq k+1 \, .
\end{eqnarray}
Usual techniques \cite {FGR,FGR2} allow to find the unknowns $\tilde
b_{2n-1}^{(k)},\tilde b_{2n}^{(k)}$ as solutions of the two
recursive equations
\begin{eqnarray}
\tilde b_{2n}^{(k)}&=&\sum _{m=0}^{n-k}(-1)^m 2^{m+\frac {1}{2}}
\frac {\tilde b_{2n-2m+1}^{(k)}}{(2m)!} B_{2m} \, , \quad n \geq k \, , \nonumber \\
&& \label {beqs} \\
\tilde b_{2n+1}^{(k)}&=&\frac {(-1)^n}{2\pi} 2^{2k+\frac
{1}{2}-n}\frac {2^{2n-2k+1}-1}{(2n-2k+2)!}B_{2n-2k+2}+\sum
_{m=0}^{n+1-k}(-1)^m 2^{m+\frac {1}{2}} \frac {\tilde
b_{2n+2-2m}^{(k)}}{(2m)!} B_{2m} \, , \quad n \geq k \, . \nonumber
\end{eqnarray}
By comparing such equations with the corresponding equations for the
coefficients $b_N$ appearing in the asymptotic solution for
$S^{(1)}_N (g)$ we find the simple correspondence
\begin{equation}
\tilde b_N^{(k)}=\frac {(-1)^{k+1}2^k}{2\pi}b_{N-2k} \, , \quad N
\geq 2k \, .
\end{equation}
Putting all the relevant relations inside (\ref {forterm22}) and
redefining (for conciseness' sake) the indexes $l$ and $s$, we
finally find the asymptotic expansions\footnote {The notation $[x]$
present in (\ref {s2pn}, \ref {s2p1n}) stands for the integer part
of the semi-integer $x$.}
\begin{eqnarray}
S_{2p}^{(n)}(g)&\doteq&\frac {2p}{\pi} \sum _{k=1}^n \sum _{l=1}^{\left [
\frac {k-1}{2}\right ]} (-1)^l \Bigl [ \sum _{s=0}^{l-1}
\left(\begin{array}{c}2l \\2s\end{array}\right) (-1)^s \sigma ^{(n-k);(2s)} \cdot \nonumber \\
&\cdot & \sum _{n'=0}^{\infty}\frac
{2^{l-s}(-1)^{n'}}{g^{2n'+2l-2s}}\frac {\Gamma (p+n'+l-s)} {\Gamma
(p-n'-l+s+1)} b_{2n'} \Bigr ] \sum_{ \{j_1,\dots,j_{k-2l} \} }
\prod_{m=1}^{k-2l} \frac{(c^{(m)})^{j_m}}{ j_m! }, \label {s2pn}
\end{eqnarray}
and
\begin{eqnarray}
S_{2p-1}^{(n)}(g)&\doteq&\frac {2p-1}{\pi} \sum _{k=1}^n \sum
_{l=1}^{\left [ \frac {k-1}{2}\right ]} (-1)^l \Bigl [ \sum
_{s=0}^{l-1} \left(\begin{array}{c}2l \\2s\end{array}\right) (-1)^s
\sigma ^{(n-k);(2s)}
 \cdot  \nonumber \\
&\cdot & \sum _{n'=0}^{\infty}\frac
{2^{l-s}(-1)^{n'}}{g^{2n'+2l-2s+1}}\frac {\Gamma (p+n'+l-s)} {\Gamma
(p-n'-l+s)} b_{2n'+1} \Bigr ] \sum_{ \{j_1,\dots,j_{k-2l} \} }
\prod_{m=1}^{k-2l} \frac{(c^{(m)})^{j_m}}{ j_m! } \, ,  \label
{s2p1n}
\end{eqnarray}
where the positive integers $j_m$ have to satisfy
\begin{equation}
\sum _{m=1}^{k-2l}j_m=2l+1 \, , \quad \sum _{m=1}^{k-2l}m j_m=k \, .
\end{equation}

We are now ready to discuss the strong coupling behaviour of the
densities of Bethe roots and their derivatives at $u=0$,
$\sigma^{(n);(s)}$, when $n\geq 1$. We have to distinguish between
the case $n=1$ and the cases $n\geq 2$.

In the case $n=1$ we have that
\begin{equation}
\sigma _0^{(1);(0)}=-4\ln 2 \, , \quad i^{-s}
\sigma^{(1);(s)}_0=d_{s}(4-2^{s+2})\Gamma (s+1)\zeta (s+1) \, , \quad s
\geq 2 \, ,
\end{equation}
for the one loop theory and
\begin{equation}
i^{-s}  \sigma^{(1);(s)}_H= d_{s} \int_0^{+\infty} dk \, \frac{k^{s}
}{\sinh k/2} \,\sum_{p=1}^\infty S^{(1)}_p (g) \, J_p(\sqrt{2} g k)
\, , \label {higher}
\end{equation}
for the higher than one loop contributions. We now insert in (\ref
{higher}) the asymptotic solution \cite {FGR} for $S_p^{(1)}(g)$,
reported also in (\ref {sm}, \ref {aeul}). Summing first over $p$
and then over $n'$ and finally integrating in $k$ we get\footnote{
The numerical analysis gives a convincing evidence that the leading correction to the densities (\ref{densh1}) is exponentially small and behaves like $ O\left ( e^{-\frac
{2 \pi g}{\sqrt {2}}} \right )$. An analytic proof of this fact is still lacking.}
\begin{eqnarray}
 \sigma^{(1);(0)}_H&=&3\ln 2 -\frac {\pi}{2} + O\left ( e^{-\frac
{2 \pi g}{\sqrt {2}}} \right )\, , \nonumber \\
\label {densh1} \\
i^{-s} \sigma^{(1);(s)}_H&=&d_{s}\left [ (2^{s+2}-4+2^{-2s}-2^{-s})\Gamma
(s+1)\zeta (s+1) -\left (\frac {\pi}{2}\right )^{s+1} |E_{s}| + O\left ( e^{-\frac
{2 \pi g}{\sqrt {2}}} \right ) \right ] \, ,
\quad s \geq 2 \, , \nonumber
\end{eqnarray}
where $E_k$ are the Euler's numbers. Adding the one loop
results, one gets the explicit formula
\begin{eqnarray}
\sigma^{(1);(0)}&=&-\ln 2 -\frac {\pi}{2} + O\left ( e^{-\frac
{2 \pi g}{\sqrt {2}}} \right )\, , \nonumber \\
\label {dens1} \\
i^{-s} \sigma^{(1);(s)}&=&d_{s}\left [ (2^{-2s}-2^{-s})\Gamma (s+1)\zeta
(s+1) -\left (\frac {\pi}{2}\right )^{s+1} |E_{s}|+ O\left ( e^{-\frac
{2 \pi g}{\sqrt {2}}} \right ) \right ]  \, , \quad s \geq
2 \, . \nonumber
\end{eqnarray}

For the case $n\geq 2$ we start from (\ref {Sndefi}) and write
\begin{equation}
i^{-s}  \sigma^{(n);(s)}= 2 d_{s} \int _{0}^{+\infty}\frac {dk}{2\pi}
k ^s
 \Bigl [\frac {\pi k}{\sinh \frac {k}{2}}S^{(n)}(k)
-\frac {e^{-\frac {k}{2}}}{\sinh \frac {k}{2}}\int  _{-\infty
}^{+\infty} \frac {dp}{2\pi} \left . \hat \sigma (p) \frac {\sin
(k-p)c} {k-p} \right |_{j^n} \Bigr] \, . \label{signs}
 \end{equation}
We then insert (\ref {s2pn}, \ref {s2p1n}) in the expansion (\ref
{snbess}) of $S^{(n)}(k)$ in series of Bessel functions and use
(\ref {jexp}) to express the integral term in the square brackets.
Summing on the indexes $p$ and $n'$ coming from (\ref {s2pn}, \ref
{s2p1n}), we end up with the previously announced strong coupling recursive
equation,
\begin{eqnarray}
i^{-s}  \sigma^{(n);(s)}&=& d_{s} \sum _{k=1}^n \sum _{l=1}^{\left [
\frac {k-1}{2}\right ]} (-1)^l \sum _{s'=0}^{l-1}
\left(\begin{array}{c}2l \\2s'\end{array}\right) (-1)^{s'} \sigma
^{(n-k);
(2s')} \cdot \nonumber \\
&\cdot & \int _{0}^{+\infty}\frac {dk}{\pi}\frac
{k^{s+2l-2s'}}{\sinh \frac {k}{2}}\left (\frac {e^{\frac
{k}{2}}}{\cosh k}-e^{-\frac {k}{2}}\right )
 \sum_{ \{j_1,\dots,j_{k-2l} \} }  \prod_{m=1}^{k-2l}
\frac{(c^{(m)})^{j_m}}{ j_m! }+ \ldots = \nonumber \\
&=& d_{s} \sum _{k=1}^n \sum _{l=1}^{\left [ \frac {k-1}{2}\right ]}
(-1)^l \sum _{s'=0}^{l-1} \left(\begin{array}{c}2l
\\2s'\end{array}\right) (-1)^{s'} \sigma ^{(n-k);
(2s')} \cdot \label{recurdens} \\
&\cdot & \frac {1}{\pi} \Bigl [ (2^{-s-2l+2s'}-2^{-2s-4l+4s'})\Gamma (s+2l-2s'+1)\zeta (s+2l-2s'+1) +  \nonumber \\
&+& \left (\frac {\pi}{2}\right )^{s+2l-2s'+1} |E_{s+2l-2s'}| \Bigr
] \sum_{ \{j_1,\dots,j_{k-2l} \} }  \prod_{m=1}^{k-2l}
\frac{(c^{(m)})^{j_m}}{ j_m! }+ \ldots  \, , \nonumber
\end{eqnarray}
where, again,
\begin{equation}
\sum _{m=1}^{k-2l}j_m=2l+1 \, , \quad \sum _{m=1}^{k-2l}m j_m=k \, .
\end{equation}
As we said before, this equation has to be solved together with
(\ref{recurcoeff}).

To summarize the results, in this section we have shown that,
similarly to the case of $S^{(1)}_{r}(g)$ \cite {FGR}, also the
system for the $\tilde S^{(k)}_{r}(g)$ can be solved in form of an
asymptotic series at large $g$.
This allowed to write for the $\sigma ^{(n');(s)}$, $n'\geq 2$, the
recursion relation (\ref{recurdens}) - which goes together with the
explicit expressions (\ref {dens1}, \ref {cSm}), coming from the
solution of the systems for $S^{(1)}_{r}(g)$ and from results
contained in Appendix A, respectively. In this (strong coupling) regime, the set of constants
$c^{(m)} $ and the set of densities $\sigma^{(n');(k)}$, $n'\geq 2$, can be
computed by solving simultaneously the recursive relations
(\ref{recurcoeff}) and (\ref{recurdens}). Putting their expressions,
together with the expressions (\ref {cSm}) for $\tilde S^{(k)}_{1}(g)$,
$\sigma^{(0);(2k)}$, and (\ref {dens1}) for $\sigma^{(1);(2k)}$, into (\ref
{forterm22}), one will get the expression for $S^{(n)}_{1}(g)$ and,
consequently, for $f_n(g)={\sqrt {2}}g S^{(n)}_{1}(g)$ at strong
coupling.

As an application of all these techniques, in the next section we
will compute explicitly the strong coupling limit of the scaling
functions $f_3(g), \ldots ,f_8(g)$.

\subsection{Examples: $f_3(g)$ to $f_8(g)$}

The previous machinery can be tested by computing the strong
coupling behaviour of $f_n(g)$, for $3\leq n \leq 8$, in order to
compare it with the available results from the $O(6)$ NLSM.
In this respect we remember that a recent proposal by Alday and Maldacena \cite {AM}, formulated on the
string side of the AdS/CFT correspondence, states that in the limit (\ref {jlimit}),
when $g\rightarrow \infty$, $j \ll g$, with
$j/m(g)$ fixed, the quantity $f(g,j)+j$ has to coincide with the energy density
of the ground state of the $O(6)$ nonlinear sigma model. When $j/m(g) \ll 1$ we are
in the nonperturbative regime of the $O(6)$ NLSM. In this case the energy density
can be computed by using Bethe Ansatz related techniques. This computation has been systematically
performed in \cite {BF}. In order to have agreement between our calculations for $f(g,j)$
and computations of \cite {BF} in the $O(6)$ NLSM, we must have that the quantities
$\Omega_n(g)$ computed in that paper have to be related to $f_n(g)$ by the relation
$f_n(g)=2^{n-1}\Omega_n(g)$. In this subsection we will check this equality,
for $3\leq n \leq 8$, by using also symbolic manipulations performed by means of a 
{\it Mathematica}$^{\textrm{\textregistered}}$ code
reported in Appendix B.

First of all, we need to know the expression of $c^{(1)},\ldots
,c^{(6)}$. From the recursion formula (\ref {recurcoeff}) we get
\begin{eqnarray}
c^{(1)} & = &-\frac {\pi}{\sigma^{(0);(0)}} \, , \\
c^{(2)} & = & \pi \frac {\sigma ^{(1);(0)}}{[\sigma ^{(0);(0)}]^2} \, , \\
c^{(3)} & = & \frac{\pi^3}{6}
\frac{\sigma^{(0);(2)}}{[\sigma^{(0);(0)}]^4} -
\pi \frac{[\sigma^{(1);(0)}]^2}{[\sigma^{(0);(0)}]^3} \, , \\
c^{(4)} & = & \pi  \frac{\sigma^{(3);(0)}}{[\sigma^{(0);(0)}]^2}-
\frac {2}{3}\pi ^3
\frac{\sigma^{(0);(2)}\sigma^{(1);(0)}}{[\sigma^{(0);(0)}]^5}+ \pi
\frac{[\sigma^{(1);(0)}]^3}{[\sigma^{(0);(0)}]^4} +\frac {1}{6} \pi
^3
\frac{\sigma^{(1);(2)}}{[\sigma^{(0);(0)}]^4} \, , \\
c^{(5)} & = & -\frac{\pi [ \sigma ^{(1);(0)}]^4}{[\sigma ^{(0);(0)}]^5}+\frac{5 \pi ^3 \sigma ^{(0);(2)} [\sigma ^{(1);(0)}]^2}{3 [\sigma^{(0);(0)}]^6}-\frac{2 \pi ^3 \sigma ^{(1);(2)} \sigma ^{(1);(0)}}{3  [\sigma^{(0);(0)}]^5}-\frac{2 \pi  \sigma ^{(3);(0)} \sigma ^{(1);(0)}}{[\sigma ^{(0);(0)}]^3}+ \,  \nonumber \\
& - & \frac{\pi ^5 [\sigma ^{(0);(2)}]^2}{12 [\sigma ^{(0);(0)}]^7}+\frac{\pi ^5 \sigma ^{(0);(4)}}{120 [\sigma ^{(0);(0)}]^6}+\frac{\pi  \sigma ^{(4);(0)}}{[\sigma ^{(0);(0)}]^2}  \, , \\
c^{(6)} & = & \frac{\pi [ \sigma ^{(1);(0)}]^5}{[\sigma ^{(0);(0)}]^6}-\frac{10 \pi ^3 \sigma ^{(0);(2)} [\sigma ^{(1);(0)}]^3}{3 [\sigma ^{(0);(0)}]^7}+\frac{5 \pi ^3 \sigma ^{(1);(2)} [\sigma ^{(1);(0)}]^2}{3 [\sigma ^{(0);(0)}]^6}+\frac{3 \pi    \sigma ^{(3);(0)} [ \sigma ^{(1);(0)}]^2}{[\sigma ^{(0);(0)}]^4}+ \nonumber  \\
&+ &\frac{7 \pi ^5 [\sigma ^{(0);(2)}]^2 \sigma ^{(1);(0)}}{12 [\sigma ^{(0);(0)}]^8}-\frac{\pi ^5 \sigma ^{(0);(4)} \sigma ^{(1);(0)}}{20 [\sigma ^{(0);(0)}]^7}-\frac{2 \pi  \sigma ^{(4);(0)} \sigma ^{(1);(0)}}{[\sigma ^{(0);(0)}]^3}-\frac{\pi ^5 \sigma ^{(0);(2)} \sigma ^{(1);(2)}}{6 [\sigma ^{(0);(0)}]^7}+  \nonumber \\
& + &\frac{\pi ^5 [\sigma ^{(1);(4)}]}{120 [\sigma
^{(0);(0)}]^6}-\frac{2 \pi ^3 \sigma ^{(0);(2)} \sigma ^{(3);(0)}}{3
[\sigma ^{(0);(0)}]^5}+\frac{\pi ^3 [\sigma ^{(3);(2)}]}{6 [\sigma
^{(0);(0)}]^4}+\frac{\pi  [\sigma ^{(5);(0)}]}{[\sigma
^{(0);(0)}]^2}.
\end{eqnarray}
Then, referring for the notations to (\ref {jexp}), we have
\begin{eqnarray}
&&2 \int_{-\infty}^\infty \frac{dp}{2 \pi} \, \left[ \Gamma_3(h-p,p)
-
\Gamma_3(p,p) \right] =\frac {1}{3}\pi ^3 \frac {h^2}{[\sigma^{ (0);(0)}]^2} \\
&&2 \int_{-\infty}^\infty \frac{dp}{2 \pi} \, \left[ \Gamma_4(h-p,p)
- \Gamma_4(p,p) \right] = -\frac {2}{3}\pi ^3
\frac {h^2 \sigma^{(1);(0)}}{[\sigma^{ (0);(0)}]^3} \\
&&2 \int_{-\infty}^\infty \frac{dp}{2 \pi} \, \left[ \Gamma_5(h-p,p)
-
\Gamma_5(p,p) \right] =  -\frac{\pi ^5 h^4}{60 [\sigma^{(0);(0)}]^4}+\frac{\pi ^3 [\sigma^{ (1);(0)}]^2 h^2}{[\sigma^{ (0);(0)}]^4}-\frac{\pi ^5 \sigma^{ (0);(2)} h^2}{15 [\sigma^{(0);(0)}]^5} \\
&&2 \int_{-\infty}^\infty \frac{dp}{2 \pi} \, \left[ \Gamma_6(h-p,p)
- \Gamma_6(p,p) \right] = \frac{\pi ^5 h^4 \sigma^{(1);(0)}}{15
[\sigma^{(0);(0)}]^5}- \frac {2}{3} \frac{\pi ^3 \sigma^{
(3);(0)}h^2}{[\sigma^{ (0);(0)}]^3}+ \frac {1}{3}\frac{\pi ^5
\sigma^{(1);(0)} \sigma^{ (0);(2)} h^2}
{[\sigma^{(0);(0)}]^6} +\nonumber \\
& - & \frac {4}{3} \frac{\pi ^3 [\sigma^{ (1);(0)}]^3 h^2}{[\sigma^{
(0);(0)}]^5}
-\frac {1}{15} \frac{\pi ^5 \sigma^{ (1);(2)} h^2}{[\sigma^{ (0);(0)}]^5} \, \\
&& 2  \int_{-\infty}^\infty \frac{dp}{2 \pi} \, \left[
\Gamma_7(h-p,p) - \Gamma_7(p,p) \right] =
\frac{\pi ^7 h^6}{2520 [\sigma ^{(0);(0)}]^6}-\frac{\pi ^5 [\sigma ^{(1);(0)}]^2 h^4}{6 [\sigma ^{(0);(0)}]^6}+\frac{\pi ^7 \sigma ^{(0);(2)} h^4}{126 [\sigma ^{(0);(0)}]^7}+ \nonumber \\
&+&
\frac{5 \pi ^3 [\sigma^{ (1);(0)}]^4 h^2}{3 [\sigma ^{(0);(0)}]^6}+\frac{\pi ^7 [\sigma ^{(0);(2)}]^2 h^2}{36 [\sigma ^{(0);(0)}]^8}-\frac{\pi ^5 \sigma ^{(0);(2)} [\sigma ^{(1);(0)}]^2 h^2}{[\sigma ^{(0);(0)}]^7}-\frac{\pi ^7  \sigma^{ (0);(4)} h^2}{420 [\sigma ^{(0);(0)}]^7}+\frac{\pi ^5 [\sigma ^{(1);(0)}] \sigma ^{(1);(2)} h^2}{3 [\sigma ^{(0);(0)}]^6}+ \nonumber \\
&+ &\frac{2 \pi ^3 [\sigma ^{(1);(0)}] \sigma ^{(3);(0)}
h^2}{[\sigma ^{(0);(0)}]^4}-\frac{2
   \pi ^3 \sigma ^{(4);(0)} h^2}{3 [\sigma ^{(0);(0)}]^3} \, \\
&& 2  \int_{-\infty}^\infty \frac{dp}{2 \pi} \, \left[
\Gamma_8(h-p,p) -
\Gamma_8(p,p) \right] =  -\frac{\pi ^7 [\sigma ^{(1);(0)}] h^6}{420 [\sigma ^{(0);(0)}]^7}+\frac{\pi ^5 [\sigma ^{(1);(0)}]^3 h^4}{3 [\sigma ^{(0);(0)}]^7}-\frac{\pi ^7 \sigma ^{(0);(2)} [\sigma ^{(1);(0)}] h^4}{18 [\sigma ^{(0);(0)}]^8} + \nonumber \\
& + & \frac{\pi ^7 \sigma ^{(1);(2)} h^4}{126 [\sigma ^{(0);(0)}]^7}+\frac{\pi ^5 \sigma ^{(3);(0)} h^4}{15 [\sigma ^{(0);(0)}]^5}-\frac{2 \pi ^3 [\sigma ^{(1);(0)}]^5 h^2}{[\sigma ^{(0);(0)}]^7}+\frac{7 \pi ^5 \sigma ^{(0);(2)} [\sigma ^{(1);(0)}]^3 h^2}{3 [\sigma ^{(0);(0)}]^8}-\frac{2 \pi ^7 [\sigma ^{(0);(2)}]^2 [\sigma ^{(1);(0)}] h^2}{9 [\sigma ^{(0);(0)}]^9}+ \nonumber \\
& + &\frac{\pi ^7 \sigma ^{(0);(4)} [\sigma ^{(1);(0)}] h^2}{60 [\sigma ^{(0);(0)}]^8}-\frac{\pi ^5 [\sigma ^{(1);(0)}]^2 \sigma ^{(1);(2)} h^2}{[\sigma ^{(0);(0)}]^7}+\frac{\pi ^7 \sigma ^{(0);(2)} \sigma ^{(1);(2)} h^2}{18 [\sigma ^{(0);(0)}]^8}-\frac{\pi ^7 [\sigma ^{(1);(4)}] h^2}{420 [\sigma ^{(0);(0)}]^7} -  \nonumber \\
& - &  \frac{4 \pi ^3 [\sigma ^{(1);(0)}]^2 \sigma ^{(3);(0)}
h^2}{[\sigma ^{(0);(0)}]^5} + \frac{\pi ^5 \sigma ^{(0);(2)} \sigma
^{(3);(0)}
   h^2}{3 [\sigma ^{(0);(0)}]^6}-\frac{\pi ^5 [\sigma ^{(3);(2)}] h^2}{15 [\sigma ^{(0);(0)}]^5}+\frac{2 \pi ^3 [\sigma ^{(1);(0)}] \sigma ^{(4);(0)} h^2}{[\sigma ^{(0);(0)}]^4}- \nonumber \\
&-&\frac{2 \pi ^3 [\sigma ^{(5);(0)}]
   h^2}{3 [\sigma ^{(0);(0)}]^3} .
\end{eqnarray}
This implies that for $f_3(g),\ldots , f_8(g)$ we can give the exact
({\it i.e.} valid $\forall \, g$) expressions:
\begin{eqnarray}
\frac {f_3(g)}{2{\sqrt {2}g}}&=&\frac {1}{6}\pi ^3 \frac {1}
{[\sigma^{ (0);(0)}]^2} \tilde S_1^{(1)}(g) \, , \label{f3all}  \\
\frac {f_4(g)}{2{\sqrt {2}g}}&=& -\frac {1}{3}\pi ^3
\frac {\sigma^{(1);(0)}}{[\sigma^{ (0);(0)}]^3} \tilde S_1^{(1)}(g)\, , \label{f4all} \\
\frac {f_5(g)}{2{\sqrt {2}g}}&=& -\frac{\pi ^5 }{120 [\sigma^{(0);(0)}]^4} \tilde S_1^{(2)}(g)+\frac {1}{2} \frac{\pi ^3 [\sigma^{ (1);(0)}]^2 }{[\sigma^{ (0);(0)}]^4} \tilde S_1^{(1)}(g)-\frac{\pi ^5 \sigma^{ (0);(2)} }{30 [\sigma^{(0);(0)}]^5} \tilde S_1^{(1)}(g) \, , \label{f5all} \\
\frac {f_6(g)}{2{\sqrt {2}g}}&=& \frac{\pi ^5 \sigma^{(1);(0)}}{30
[\sigma^{(0);(0)}]^5} \tilde S_1^{(2)}(g)+ \Biggl [ - \frac {1}{3}
\frac{\pi ^3 \sigma^{ (3);(0)}}{[\sigma^{ (0);(0)}]^3}+\label{f6all}
\\
&+&\frac {1}{6}\frac{\pi ^5  \sigma^{(1);(0)} \sigma^{ (0);(2)} }
{[\sigma^{(0);(0)}]^6}- \frac {2}{3} \frac{\pi ^3 [\sigma^{
(1);(0)}]^3 }{[\sigma^{ (0);(0)}]^5} -\frac {1}{30} \frac{\pi ^5
\sigma^{ (1);(2)} }{[\sigma^{ (0);(0)}]^5}\Biggr ]
 \tilde S_1^{(1)}(g) \, ,  \nonumber  \\
\frac {f_7(g)}{2{\sqrt {2}g}}&=&
\frac{\pi ^7 }{5040 [\sigma ^{(0);(0)}]^6}   \tilde S_1^{(3)}(g) +\Biggl [-\frac{\pi ^5 [\sigma ^{(1);(0)}]^2 }{12 [\sigma ^{(0);(0)}]^6}+\frac{\pi ^7 \sigma ^{(0);(2)}}{252 [\sigma ^{(0);(0)}]^7}\Biggr ]  \tilde S_1^{(2)}(g)+ \label{f7all} \\
&+& \Biggl [ \frac{5 \pi ^3 [\sigma ^{(1);(0)}]^4}{6 [\sigma
^{(0);(0)}]^6}+\frac{\pi ^7 [\sigma ^{(0);(2)}]^2 }{72 [\sigma
^{(0);(0)}]^8} - \frac{\pi ^5 \sigma ^{(0);(2)} [\sigma
^{(1);(0)}]^2}{2 [\sigma ^{(0);(0)}]^7} -
\frac{\pi ^7 \sigma ^{(0);(4)}}{840 [\sigma ^{(0);(0)}]^7}+ \nonumber \\
&+ & \frac{\pi ^5 [\sigma ^{(1);(0)}] \sigma ^{(1);(2)} }{6 [\sigma
^{(0);(0)}]^6}+ \frac{ \pi ^3 [\sigma ^{(1);(0)}] \sigma ^{(3);(0)}
}{[\sigma ^{(0);(0)}]^4}-\frac{
   \pi ^3 \sigma ^{(4);(0)} }{3 [\sigma ^{(0);(0)}]^3} \Biggr ]  \tilde S_1^{(1)}(g)\, ,  \nonumber \\
\frac {f_8(g)}{2{\sqrt {2}g}}&=&  -\frac{\pi ^7 [\sigma ^{(1);(0)}] }{840 [\sigma ^{(0);(0)}]^7}  \tilde S_1^{(3)}(g)  + \Biggl [  \frac{\pi ^5 [\sigma ^{(1);(0)}]^3}{6 [\sigma ^{(0);(0)}]^7}-\frac{\pi ^7 \sigma ^{(0);(2)} [\sigma ^{(1);(0)}] }{36 [\sigma ^{(0);(0)}]^8} +  \frac{\pi ^7 \sigma ^{(1);(2)} }{252 [\sigma ^{(0);(0)}]^7}  + \label{f8all} \\
& + &\frac{\pi ^5 \sigma ^{(3);(0)} }{30 [\sigma ^{(0);(0)}]^5} \Biggr ]  \tilde S_1^{(2)}(g) + \Biggl [  -\frac{ \pi ^3 [\sigma ^{(1);(0)}]^5 }{[\sigma ^{(0);(0)}]^7}+\frac{7 \pi ^5 \sigma ^{(0);(2)} [\sigma ^{(1);(0)}]^3}{6 [\sigma ^{(0);(0)}]^8}-\frac{ \pi ^7 [\sigma ^{(0);(2)}]^2 [\sigma ^{(1);(0)}] }{9 [\sigma ^{(0);(0)}]^9}+ \nonumber \\
& + &\frac{\pi ^7 \sigma ^{(0);(4)} [\sigma ^{(1);(0)}] }{120 [\sigma ^{(0);(0)}]^8}-\frac{\pi ^5 [\sigma ^{(1);(0)}]^2 \sigma ^{(1);(2)} }{2[\sigma ^{(0);(0)}]^7}+\frac{\pi ^7 \sigma ^{(0);(2)} \sigma ^{(1);(2)}}{36 [\sigma ^{(0);(0)}]^8} -  \frac{2 \pi ^3 [\sigma ^{(1);(0)}]^2 \sigma ^{(3);(0)} }{[\sigma ^{(0);(0)}]^5}  \nonumber \\
& + & \frac{\pi ^5 \sigma ^{(0);(2)} \sigma ^{(3);(0)}
   }{6 [\sigma ^{(0);(0)}]^6}-\frac{\pi ^5 [\sigma ^{(3);(2)}] }{30 [\sigma ^{(0);(0)}]^5}+\frac{ \pi ^3 [\sigma ^{(1);(0)}] \sigma ^{(4);(0)} }{[\sigma ^{(0);(0)}]^4}-\frac{\pi ^7 [\sigma ^{(1);(4)}] }{840 [\sigma ^{(0);(0)}]^7}-\frac{ \pi ^3 [\sigma ^{(5);(0)}]
   }{3 [\sigma ^{(0);(0)}]^3}     \Biggr ]  \tilde S_1^{(1)}(g) \, , \nonumber
\end{eqnarray}
In the limit $g \rightarrow \infty$, the quantity $\sigma
^{(n);(s)}$ for $n\geq 1$ can be computed for $n=1$ by using (\ref {dens1}) and
for $n\geq 2$ by solving together the recursive equations (\ref {recurdens}) and
(\ref {recurcoeff}). In particular we have
\begin{equation}
\sigma^{ (1);(0)}=-\ln 2 - \frac {\pi}{2} + O\left ( e^{-\frac
{2 \pi g}{\sqrt {2}}} \right ) \, , \quad \sigma^{
(1);(2)}=\frac {3 \zeta (3)+\pi ^3}{8} +O\left ( e^{-\frac
{2 \pi g}{\sqrt {2}}} \right ) \, , \quad \sigma^{ (3);(0)}=
\frac{\pi ^2}{6[\sigma ^{(0);(0)}]^2} \sigma^{ (1);(2)}\, , \nonumber
\end{equation}
\begin{equation}
\sigma^{ (1);(4)}=-\frac{5}{32} (\pi^5 + 9 \zeta(5))+O\left ( e^{-\frac
{2 \pi g}{\sqrt {2}}} \right ) \, , \quad
\sigma^{ (3);(2)}= \frac{\pi ^2}{6[\sigma ^{(0);(0)}]^2} \sigma^{
(1);(4)} \, , \quad \sigma^{ (4);(0)}=-\pi ^2\frac {\sigma
^{(1);(0)} \sigma ^{(1);(2)}}{3[\sigma ^{(0);(0)}]^3}\, , \nonumber
\end{equation}
\begin{equation}
\sigma^{ (5);(0)}=\frac{\pi ^2 \sigma ^{(1);(2)} [\sigma
^{(1);(0)}]^2}{2 [\sigma ^{(0);(0)}]^4}-\frac{\pi ^4 \sigma
^{(0);(2)} \sigma ^{(1);(2)}}{30 [\sigma ^{(0);(0)}]^5}+\frac{\pi ^4
[\sigma ^{(1);(4)}]}{120 [\sigma ^{(0);(0)}]^4} \, ,
\end{equation}
where for $\sigma ^{(0);(2k)}$ we have to use the strong coupling expressions (\ref {cSm}).
These formul{\ae}, together with the values at strong
coupling of $\sigma ^{(0);(s)}$ and $\tilde S_1^{(k)}(g)$ (\ref
{cSm}), allow to obtain for $f_3(g),\ldots , f_8(g)$ the following
(leading) values as $g\rightarrow \infty$,
\begin{eqnarray}
f_3(g)&=& \frac {\pi ^2}{24m(g)} + O\left ( e^{-\frac
{\pi g}{\sqrt {2}}} \right ) \, , \label {f3} \\
f_4(g)&=& -\frac {\pi ^2}{12 [m(g)]^2} {\cal S}_1 + O(1) \, ,  \label {f4} \\
f_5(g)&=&-\frac {\pi ^4}{640 [m(g)]^3}+\frac {\pi ^2}{8[m(g)]^3}[{\cal S}_1]^2 + O\left ( e^{\frac
{\pi g}{\sqrt {2}}} \right ) \, ,  \label {f5} \\
f_6(g)&=&\frac {\pi ^4}{[m(g)]^4} \left (\frac {{\cal
S}_3}{90}-\frac
{[{\cal S}_1]^3}{6 \pi^2}+\frac {{\cal S}_1}{120} \right )+ O\left ( e^{\frac
{2 \pi g}{\sqrt {2}}} \right ) \, ,  \label {f6} \\
f_7(g)&=&\frac {\pi ^6}{7182 [m(g)]^5}-\frac{5 \pi ^4 [{\cal S}_1]^2
} {192[m(g)]^5}+\frac {5 \pi ^2 [{\cal S}_1]^4 }{24[m(g)]^5} -
\frac{\pi ^4 {\cal S}_1{\cal S}_3}{18[m(g)]^5} + \ldots \, ,  \label {f7} \\
f_8(g)&=&-\frac{ \pi ^6 {\cal S}_1 }{840 [m(g)]^6}+\frac { \pi ^4
[{\cal S}_1]^3 }{16[m(g)]^6}  -\frac { \pi ^2 [{\cal S}_1]^5
}{4[m(g)]^6}-\frac{ \pi ^6 {\cal S}_3 }{560[m(g)]^6}+ \frac{\pi ^4
{\cal S}_1^2 {\cal S}_3}{6 [m(g)]^6} -\nonumber \\
&-& \frac{ \pi ^6 {\cal S}_5 }{280
[m(g)]^6} + \ldots \, ,  \label {f8}
\end{eqnarray}
where we used the compact notations:
\begin{equation}
{\cal S}_{2s+1}=\frac {1}{\pi ^{2s+1}}\sum _{n=0}^{\infty} (-1)^n
\left [ \frac {1}{\left (n+\frac {1}{2} \right )^{2s+1}}+ \frac
{1}{(n+1)^{2s+1}} \right ] \, . \label {calS}
\end{equation}
For instance, we have
\begin{equation}
{\cal S}_1=\frac {1}{\pi}\ln 2 +\frac {1}{2} \, , \quad {\cal
S}_3=\frac {1}{4\pi ^3}[3\zeta (3)+\pi ^3] \, , \quad {\cal S}_5=
\frac {5}{48\pi ^5}[9\zeta (5)+\pi ^5] \, .
\end{equation}
After a lengthy but straightforward calculation it is possible to
show that such expressions agree with the
corresponding formul{\ae}\footnote{In order to perform such a check, we have explicitly calculated $\Omega_n$ up to $n=8$ according to the expressions found in \cite{BF}.} computed in the framework of the $O(6)$
NLSM, {\it i.e.} the coefficients $2^{n-1}\Omega_n(g)$ given by the general
formulae of \cite{BF}.

\medskip

It emerges from our analysis that at the leading strong coupling order the generalised scaling functions $f_n(g)$ (and then $f(g,j)$) are all dominated by the $O(6)$ NLSM energy density contribution. This implies that they are all given by a suitable power of the (unique) NLSM mass-gap $m(g)$, {\it i.e.} $f_n(g) = a_n [m(g)]^{2-n}+\dots$ where the $a_n$ can be computed within the NLSM \cite{BF} or the formul\ae \, (\ref{f3})-(\ref{f8}). This fact motivates the introduction of the following quantities or ``masses"
\begin{equation}
m_n(g) \equiv \left( \frac{a_n}{f_n(g)} \right)^{\frac{1}{n-2}} \, ,
\end{equation}
which all tend to the unique NLSM mass-gap $m(g)$. Beyond the leading order and indeed for all $g$, the generalised scaling functions $f_n(g)$ can be deduced by putting together all the relevant results for ${\tilde S}^{(k)}_1(g)$ and $\sigma^{(n);(s)}$, and consequently the ``masses" above expand as
\begin{equation}
m_n(g) = m(g) + p_n(g) \, g^{\delta_n}\, e^{-\frac{3 \pi g}{\sqrt
2}} + \dots \, ,
\end{equation}
where the $p_n (g)=p^0_n +O(1/g)$ can be expressed as (asymptotic) expansions in the variable $1/g$, the $\delta_n$ are some constants, and the dots stand for higher order non-analytic corrections.
These expansions for the masses $m_n(g)$ are of particular interest, because at the order
$O\left (e^{-\frac{ \pi g}{\sqrt 2}} \right )$ all these $m_n(g)$ reduce to the unique NLSM mass-gap:
this is a convergence phenomenon that agrees with the simplification
of the string dynamics observed in  \cite{AM}. In other words, all these masses converge to one, the mass gap of the $O(6)$ NLSM, because, as proposed in \cite{AM} within the dual string description, in the scaling (\ref {jlimit}) the strong coupling limit $g \gg j$ of the quantity $f(g,j)+j$ must coincide with the energy density of the $O(6)$ NLSM. Moreover, the unique mass parameter in the $O(6)$ NLSM theory is the mass gap $m(g)$.

Furthermore, it is interesting to notice that all the $m_n(g)$ share the same next-to-leading exponential decay of order $O( e^{-\frac{3 \pi g}{\sqrt 2}})$, being the difference between these masses encapsulated in the functions $p_n(g)$. In principle, the exponents $\delta_n$ would be different as well. Yet, in the next section we will perform a numerical analysis of the next to leading order corrections for various densities and reduced scaling functions and will gain some evidence for the equality of these exponents.

\subsection{Numerical evaluation of the next to leading corrections}

In order to check the results presented in the previous sections we can
numerically estimate (see Appendix C for details on the numerics) the deviations
from the leading behaviour
at strong coupling for various quantities:
\begin{equation}
\alpha_1(g)= f_1(g) + 1 \, , \quad \alpha_2(g)= - \frac{\sigma
^{(0);(0)}}{\pi} \, , \quad \alpha_3(g)= - 4 \frac{\sigma
^{(0);(2)}}{\pi^3}  \, ,
\end{equation}
\begin{equation}
\alpha_4(g)=  \frac{8}{\pi} [\sqrt2 g \tilde{S}^{(1)}_1]\, , \quad
\alpha_5(g)=- \frac{16}{\pi^3}  [\sqrt2 g  \tilde{S}^{(2)}_1]\, .
\end{equation}
As we showed in the Appendix A, all of them, indeed, at strong coupling approach
the $O(6)$ mass gap $m(g)$, up to terms $O(e^{-\frac{3 \pi g}{\sqrt 2}})$, {\it i.e.}
\begin{equation}
\alpha_i(g) = m(g) + \epsilon_i \, g^{\gamma_i}\, e^{-\frac{3 \pi g}{\sqrt
2}} + \dots
\end{equation}

The first step of the numerical analysis concerns the leading term $m(g)$.
In particular we are able to give a quite precise estimate of the coefficients $k_1$, $k_2$ appearing in (\ref {mgap}):
\begin{equation}
m(g)   =  k \, g^{1/4} \left(1+ \frac{k_1}{g}+  \frac{k_2}{g^2}+ \dots \right) e^{-\frac{\pi g}{\sqrt
2}}, \ \ \ \ \  k= \frac{2^{5/8} \pi^{1/4}}{ \Gamma(5/4)}.
\end{equation}

The analysis of the data at our disposal gives access to the
quantities $k_1$, $k_2$, whose best fit estimates are
\begin{equation}
k_1 = -0.0164\pm 0.0005, \ \ \ \ \ k_2 = -0.0026\pm 0.0004.
\end{equation}
As a by product, we are also able to check that the previous estimate is the same for all the $\alpha_i$'s (within the error bars).
As matter of facts, all the corrections in powers of $1/g$ ought to be the same for all the $\alpha_i(g)$, as shown (analytically) in appendix A.

The next step is the evaluation study of the next-to-leading terms $O(e^{-\frac{3 \pi g}{\sqrt 2}})$. The exponentially small nature of such contributions forces us to study the differences  $\Delta_{ij}(g)=\alpha_i(g) - \alpha_j(g)$,
$i<j$, in order to get rid of the leading term which would overshadow the sub-leading terms. With the usual best fit procedure, we have been able to verify that
all the $\Delta_{ij}(g)$ actually share the same pre-exponential behaviour, taking the following form
\begin{equation}
\Delta_{ij}(g)  = \delta_{i,j} k^3 g^{-1/4}e^{-\frac{3 \pi g}{\sqrt 2}} + \dots
\end{equation}
but the amplitudes $ \delta_{i,j} $ turn out to be different, reflecting the fact that the $\alpha_i$ are leaving the $O(6)$ limit following different trajectories. We put a  particular care in the check of the uniqueness of the pre-exponential factor, because such a fact strongly suggests the uniqueness of the exponents
$\gamma_i$ for all the $\alpha_i(g)$ considered here, i.e $\gamma_i = -1/4$, $\forall \, i$.

As a consistency check upon the numerical amplitudes
$\delta_{i,j}$ we verified numerically that the following identity
\begin{equation}
\delta_{i,j} + \delta_{j,k} = \delta_{i,k}
\end{equation}
actually holds for all $i,j,k$. We found that it is verified within the
numerical precision. The amplitudes are collected in table \ref{tab1}.
\begin{table}
\begin{center}
\begin{tabular}{|c|c|c|c|c|c|}\hline $\delta_{i,j}$ & 1 & 2 & 3 & 4 & 5 \\\hline 1 & 0 & -0.82(1) & -5.72(3) & 1.64(2) & 16.3(2) \\\hline 2 & 0.82(1) & 0 & -4.96(2) & 2.46(2) & 17.2(2) \\\hline 3 & 5.72(3) & 4.96(2) & 0 & 7.35(2) & 22.04(2) \\\hline 4 & -1.64(2) & -2.46(2) & -7.35(2) & 0 & 14.7(2) \\\hline 5 & -16.3(2) & -17.2(2) & -22.04(2) & -14.7(2) & 0 \\\hline \end{tabular}
\caption{Values of the amplitudes $\delta_{i,j}$. The index $i$
runs along the rows, and the index $j$ runs along the columns.}
\label{tab1}
\end{center}
\end{table}

\section{Summary and outlook}
\setcounter{equation}{0}

The aim of this article was the study of the functions $f_n(g)$
appearing in the expansion
\be
\gamma (g,s,L)= \ln s \sum _{n=0}^{\infty} f_n(g) j^n + \ldots \, ,
\ee
of the lowest anomalous dimension of twist operators of ${\cal N}=4$ SYM for fixed $j$ in the limit (\ref {jlimit}). Much help comes from the extension (\ref {egs2}) of the Kotikov-Lipatov relation \cite {KL}
\begin{equation}
\gamma (g,s,L)=\frac {1}{\pi} \lim _{k\rightarrow 0}\hat \sigma_H(k) + \dots  \, ,
\label{egs2-bis}
\end{equation}
equating the leading $\ln s$ contribution of the anomalous dimension to the Fourier transform of the higher than one loop density of roots and holes in zero, $\hat \sigma _H(0)$. The function $\hat \sigma _H(k)$ satisfies the equation (\ref {sigmaeq2}), supplemented by the one-loop equation (\ref {hatsigma0}) and by conditions (\ref {normale}) on the one-loop and all-loops separators, called $c_0$ and $c$, respectively. Usefully, the crucial quantities $c$ and $\hat \sigma _H(k)$ expand in powers of $j$:
\begin{equation}
c=\sum _{n=1}^{\infty} c^{(n)}j^n +\ldots \, , \quad \hat \sigma _H(k)=[\sum _{n=0}^{\infty} \hat \sigma _H^{(n)}(k)j^n]\ln s
+\ldots \, .
\end{equation}
As a consequence, the auxiliary function $S(k)$, defined in (\ref {Sndefi}), enjoys the same kind of expansion, (\ref {defs}). Slicing (\ref {sigmaeq2}) in powers of $j$ furnishes equation (\ref {snequ}) for all the components $S^{(n)}(k)$. Upon introducing the Neumann modes $S^{(n)}_r(g)$ (\ref {snbess}),  we constrain them by the linear infinite system (\ref {Seqn}),
\begin{eqnarray}
S^{(n)}_{2p}(g)&=&A_{2p}^{(n)}(g)-4p\sum _{m=1}^{\infty}Z_{2p,2m}(g)S^{(n)}_{2m}(g)+ 4p\sum _{m=1}^{\infty}Z_{2p,2m-1}(g)S^{(n)}_{2m-1}(g) \, , \nonumber \\
S^{(n)}_{2p-1}(g)&=&A_{2p-1}^{(n)}(g)-2(2p-1)\sum
_{m=1}^{\infty}Z_{2p-1,2m}(g)
S^{(n)}_{2m}(g)- \\
&-& 2(2p-1)\sum _{m=1}^{\infty}Z_{2p-1,2m-1}(g)S^{(n)}_{2m-1}(g)
\nonumber \, ,
\end{eqnarray}
where the infinite matrix $Z_{n,m}(g)$ is given in (\ref {Zetaint}) and the 'forcing terms'
$A_{r}^{(n)}(g)$ in (\ref {forcrn}). Crucially, (\ref{egs2-bis}) entails how easily the first mode gives the $n$-th generalised scaling function
\be
f_n(g)= {\sqrt {2}}g S^{(n)}_1(g) \ .
\ee
Manipulations in section 4 show that the solution to the above system
can be expressed as in (\ref {forterm22}). All the ingredients of this expression are detailed, in the same section, as stemming out from two sources: some ingredients ($\sigma ^{(0);(s)}$ (\ref {intdens}), $\tilde S_1^{(k)}(g)$ (\ref {intrid}), $S_1^{(1)}(g)$ (\ref {intone})) from the
solution of the BES equation, the others ($c^{(m)}, 1\leq m\leq n-2$ \, ; \ $\sigma ^{(n');(s)}$, $1\leq n'\leq n-3$) from (\ref {recurcoeff}) and from the solutions to the systems (\ref {Seq2}) and (\ref {redSeqn2}) for $S^{(1)}_{r}(g)$ and for the reduced coefficients $\tilde S_r^{(k)}(g)$, respectively. Detailed inspection of (\ref {recurcoeff}) reveals that $c^{(m)}$ depends on $\sigma ^{(n');(s)}$, with $n'\leq m-1$.
This means that $S^{(n)}_r(g)$ (and, in particular, $f_n(g)= {\sqrt {2}}g S^{(n)}_1(g)$) depends on data coming from the BES equation and from the knowledge of $S^{(n')}_r(g)$, with $n' \leq n-3$.
This implies that a recursive procedure for the determination of the $f_n(g)$ has been eventually set down. We explicitly discuss and solve this recursive procedure at large $g$ (section 5). An ingredient (and result in itself) is the asymptotic solution (\ref {s2pn}, \ref {s2p1n}) to the
'reduced' system (\ref {redSeqn2}). Together with results concerning the BES equation (reported in Appendix A) and $S^{(1)}_{r}(g)$ \cite {FGR}, this eventually has allowed us to compute
the leading (non-perturbative) orders of the generalised scaling functions; for definiteness' sake we constrained ourselves to the first eight ones (\ref {f3}-\ref {f8}).

A leading strong coupling $g\gg j$, our results match the simple calculations in the thermodynamic limit framework of the $O(6)$ nonlinear sigma model  \cite {BF}, thus confirming the Alday-Maldacena proposal \cite {AM} on the presence of the $O(6)$ nonlinear sigma model in the $sl(2)$ sector of ${\cal N}=4$ SYM \cite{BK}. Eventually, we have also detailed the deviations of the exact scaling functions $f_n(g)$ from their $O(6)$ values. For this purpose, we have found useful to parametrise the various $f_n(g)$ by quantities ('masses') all converging to the  $O(6)$ NLSM mass gap
$m(g)$ at leading order and we have computed their different corrections. This was also better illustrated by numerical evaluations of the subleading corrections
to (\ref {f3}-\ref {f8}) (Subsection 5.2).

\medskip

For what concerns future work, several directions are possible.
First, as we stated in section 2, the equations (\ref {hatsigma0},
\ref {sigmaeq2}) are suitable for the study of the subleading
correction $f^{(0)}(g,j)$ to the anomalous dimension (\ref {corrsud})
(in the regime (\ref {jlimit})). This will be the
subject of a future publication.

Then, one has to say that in the $sl(2)$
sector of ${\cal N}=4$ SYM other regimes -- {\it e.g.} large $j$ both at strong and at weak coupling \cite {CK, BEC} -- are relevant for comparisons with pure string theory results. In this respect, the
limit $s, L \rightarrow \infty$, $g \rightarrow \infty$, $l=L/(g \ln
s)$ fixed - the so-called 'semiclassical scaling limit' - has been
widely studied \cite {CK}. Application of our equations and techniques to this
case is a possible future direction of investigation.

Finally, one has to mention the new line of research related to the
recently discovered duality between ${\cal N}=6$ super Chern-Simons
(SCS) theory with $U(N)\times U(N)$ gauge group at level $k$ and
superstring theory in the $\text{AdS}_4\times\text{CP}^3$
background, when $N$ is large and the 't Hooft coupling $\lambda
=N/k$ is kept fixed \cite {ABJM}. Integrability on the gauge side
\cite {MZ2,GSY} and on the string side of the duality \cite {AF} was
shown. Bethe Ansatz-like equations were proposed \cite {GV} for the
SCS theory and tested in various ways \cite {AVGHO}. It could be
surely of interest to apply the techniques discussed in this paper
also to this new field of activity.

\vspace{0.5cm}

{\bf Acknowledgements} We thank F. Buccheri, D. Bombardelli, F.
Ravanini for discussions and suggestions. We acknowledge the INFN
grant ''Iniziativa specifica PI14'', the international agreement
INFN-MEC-2008 and the italian 'PRIN' for travel financial support.
The work of P.G. is partially supported by MEC-FEDER (grant
FPA2005-00188), by the Spanish Consolider-Ingenio 2010 Programme
CPAN (CSD2007-00042) and by Xunta de Galicia (Conseller\'\i a de
Educaci\'on and grant PGIDIT06PXIB296182PR).

\appendix

\section{Non-analytic terms at strong coupling}
\setcounter{equation}{0}

The aim of this subsection is the explicit computation at large $g$
of the leading non-analytic \footnote{These terms are not taken into account by the asymptotic expansion, because of their exponential nature (cf. section 5). In this sense they are also called non-perturbative or non-asymptotic.} contributions to the equations ~(\ref{intrid}) by using the
techniques developed in \cite{BK}.

In particular, we will calculate the large $g$ behaviour of the integrals,
\begin{eqnarray}
{\mathcal B}_{2k-1}(g) & = & \int_0^{+\infty} dt
\frac{t^{2k-1}}{\sinh t/2} \left[\sigma^{(0)}_+(\sqrt{2} g t)  -
\sigma^{(0)}_-(\sqrt{2} g t)
\right]  \, ,  \nonumber \\
{\mathcal C}_{2 k}(g) & = & \int_0^{+\infty} dt \, \frac{t^{2k}
}{\sinh t/2} \, \left[\sigma^{(0)}_+(\sqrt{2} g t)  +
\sigma^{(0)}_-(\sqrt{2} g t) \right] \nonumber \, .
\end{eqnarray}
First of all, we make use of the {\it BKK transformation} \cite{BKK,BK},
\begin{eqnarray}
2 \, \sigma^{(0)}_{\pm}(t) = \left( 1-\frac{1}{\cosh
\frac{t}{\sqrt{2} g}}  \right) \Sigma_{\pm}^{(0)} (t) \pm \tanh
\frac{t}{\sqrt{2} g} \Sigma_{\mp}^{(0)} (t) \, , \label {bkk}
\end{eqnarray}
in order to rewrite the integrals as
\begin{eqnarray}
{\mathcal B}_{2k-1}(g) & = & -  \int_0^{+\infty} \frac {dt}{t}  \,
\left( \frac{t}{\sqrt{2} g} \right)^{2k} \left[\frac{\sinh
\frac{t}{2 \sqrt{2} g} }{\cosh  \frac{t}{\sqrt{2} g}}
(\Sigma^{(0)}_-(t)  - \Sigma^{(0)}_+( t) )  -  \frac{\cosh
\frac{t}{2 \sqrt{2} g} }{\cosh  \frac{t}{\sqrt{2} g}}
(\Sigma^{(0)}_-(t)  + \Sigma^{(0)}_+( t) ) \right]
\nonumber \\
{\mathcal C}_{2 k}(g) & = &  \int_0^{+\infty} \frac {dt}{t}  \,
\left( \frac{t}{\sqrt{2} g} \right)^{2k} \left[\frac{\cosh
\frac{t}{2 \sqrt{2} g} }{\cosh  \frac{t}{\sqrt{2} g}}
(\Sigma^{(0)}_-(t)  - \Sigma^{(0)}_+( t) )  +  \frac{\sinh
\frac{t}{2 \sqrt{2} g} }{\cosh  \frac{t}{\sqrt{2} g}}
(\Sigma^{(0)}_-(t)  + \Sigma^{(0)}_+( t) ) \right]   . \nonumber
\end{eqnarray}
The BES equation can be rewritten in terms of the functions
$\Sigma_\pm^{(0)}$ \cite{BK}, with $|u|<1$
\begin{eqnarray}
&& \int_0^{+\infty} dt \, \sin(ut) [ \Sigma^{(0)}_-(t)  +
\Sigma^{(0)}_+( t) ] = 0 \, ,
\nonumber \\
&& \label{constr} \\
&& \int_0^{+\infty} dt \, \cos(ut) [ \Sigma^{(0)}_-(t)  -
\Sigma^{(0)}_+( t) ] = 2(2 \sqrt{2} g) \nonumber
\end{eqnarray}
and the ratios of hyperbolic functions admit a useful integral
representation
\begin{eqnarray}
t^{2k-1} \frac{\sinh \frac{t}{2 \sqrt{2} g} }{\cosh
\frac{t}{\sqrt{2} g}} & = & (-1)^{k+1} \, g \,
\int_{-\infty}^{+\infty} du\, \cos(ut) \, \frac{d^{2k-1}}{du^{2k-1}}
\left[ \frac{\sinh\frac{g \pi u}{ \sqrt{2} } }{\cosh \sqrt{2} g \pi
u} \right]\, ,
\nonumber \\
t^{2k-1} \frac{\cosh \frac{t}{2 \sqrt{2} g} }{\cosh
\frac{t}{\sqrt{2} g}} & = & (-1)^{k} \, g \,
\int_{-\infty}^{+\infty} du \, \sin(ut) \,
\frac{d^{2k-1}}{du^{2k-1}} \left[ \frac{\cosh \frac{g \pi u}{
\sqrt{2} } }{\cosh \sqrt{2} g \pi u} \right]
\, , \nonumber \\
t^{2k} \frac{\sinh \frac{t}{2 \sqrt{2} g} }{\cosh  \frac{t}{\sqrt{2}
g}} & = &  (-1)^{k} \, g \, \int_{-\infty}^{+\infty} du \, \sin(ut)
\, \frac{d^{2k}}{du^{2k}} \left[ \frac{\sinh \frac{g \pi u}{
\sqrt{2} } }{\cosh \sqrt{2} g \pi u} \right]\, , \nonumber \\
\nonumber t^{2k} \frac{\cosh \frac{t}{2 \sqrt{2} g} }{\cosh
\frac{t}{\sqrt{2} g}} & = &   (-1)^{k} \, g \,
\int_{-\infty}^{+\infty} du \, \cos(ut) \, \frac{d^{2k}}{du^{2k}}
\left[ \frac{\cosh \frac{g \pi u}{ \sqrt{2} } }{\cosh \sqrt{2} g \pi
u} \right]\, .
\end{eqnarray}
Plugging them into the integrals ${\mathcal B}_{2k-1}$, ${\mathcal
C}_{2k}$ we obtain
\begin{eqnarray}
{\mathcal B}_{2k-1}(g) &=&  (-1)^k g \left( \frac{1}{\sqrt{2} g}
\right)^{2k}
 \int_{-\infty}^{+\infty} du \Bigg[
 \int_0^{+\infty} dt  \, \cos(ut) \, \frac{d^{2k-1}}{du^{2k-1}} \left[ \frac{\sinh\frac{g \pi u}{ \sqrt{2} } }{\cosh \sqrt{2} g \pi u} \right]
(\Sigma^{(0)}_-(t)  - \Sigma^{(0)}_+( t) )  +
\nonumber \\
&+& \int_0^{+\infty} dt  \, \sin(ut) \, \frac{d^{2k-1}}{du^{2k-1}}
\left[ \frac{\cosh \frac{g \pi u}{ \sqrt{2} } }{\cosh \sqrt{2} g \pi
u} \right] (\Sigma^{(0)}_-(t)  + \Sigma^{(0)}_+( t) ) \Bigg]   \, ,
\nonumber \\
{\mathcal C}_{2 k}(g) & = & (-1)^k g \left( \frac{1}{\sqrt{2} g}
\right)^{2k+1} \int_{-\infty}^{+\infty} du \Bigg[ \int_0^{\infty} dt
\,  \cos(ut) \, \frac{d^{2k}}{du^{2k}} \left[ \frac{\cosh \frac{g
\pi u}{ \sqrt{2} } }{\cosh \sqrt{2} g \pi u} \right]
(\Sigma^{(0)}_-(t)  - \Sigma^{(0)}_+( t) )  + \nonumber \\ \nonumber
&+& \int_0^{+\infty} dt  \,  \sin(ut) \, \frac{d^{2k}}{du^{2k}}
\left[ \frac{\sinh \frac{g \pi u}{ \sqrt{2} } }{\cosh \sqrt{2} g \pi
u} \right] (\Sigma^{(0)}_-(t)  + \Sigma^{(0)}_+( t) ) \Bigg] \,  .
\end{eqnarray}

Let us evaluate them in the large $g$ limit. The strategy is to
split the integral over $u$ in two intervals $|u|<1$ and $|u|>1$ in
order to use the constraints (\ref{constr}). The former gives,
together with the use of such constraints:
\begin{eqnarray}
&& \int_{-1}^{1} du \, \frac{d^{2k-1}}{du^{2k-1}} \left[
\frac{\sinh\frac{g \pi u}{ \sqrt{2} } }{\cosh \sqrt{2} g \pi u}
\right]  \int_0^{+\infty} dt  \, \cos(ut)
(\Sigma^{(0)}_-(t)  - \Sigma^{(0)}_+( t) )   =  \nonumber \\
&=& 2 (2 \sqrt{2} g) \int_{-1}^{1} du \, \frac{d^{2k-1}}{du^{2k-1}}
\left[ \frac{\sinh\frac{g \pi u}{ \sqrt{2} } }{\cosh \sqrt{2} g \pi
u} \right] \, ,
\nonumber \\
&& \int_{-1}^{1} du \, \frac{d^{2k-1}}{du^{2k-1}} \left[ \frac{\cosh
\frac{g \pi u}{ \sqrt{2} } }{\cosh \sqrt{2} g \pi u} \right]
\int_0^{+\infty} dt  \, \sin(ut) \, (\Sigma^{(0)}_-(t)  +
\Sigma^{(0)}_+( t) ) = 0 \, ,
\nonumber \\
&& \int_{-1}^{1} du \, \frac{d^{2k}}{du^{2k}} \left[
\frac{\cosh\frac{g \pi u}{ \sqrt{2} } }{\cosh \sqrt{2} g \pi u}
\right]  \int_0^{+\infty} dt  \, \cos(ut)
(\Sigma^{(0)}_-(t)  - \Sigma^{(0)}_+( t) )   = \nonumber \\
&=& 2 (2 \sqrt{2} g) \int_{-1}^{1} du \, \frac{d^{2k}}{du^{2k}}
\left[ \frac{\cosh\frac{g \pi u}{ \sqrt{2} } }{\cosh \sqrt{2} g \pi
u} \right] \, ,
\nonumber \\
&& \int_{-1}^{1} du \, \frac{d^{2k}}{du^{2k}} \left[ \frac{\sinh
\frac{g \pi u}{ \sqrt{2} } }{\cosh \sqrt{2} g \pi u} \right]
\int_0^{+\infty} dt  \, \sin(ut) \, (\Sigma^{(0)}_-(t)  +
\Sigma^{(0)}_+( t) ) = 0 \, . \nonumber
\end{eqnarray}
Since we are interested in the large $g$ behaviour, we can perform
the previous integrals by rewriting them as the difference of the
integrals with support over $(-\infty,+\infty)$ and $(-\infty,-1)$,
$(1,+\infty)$, and finally taking the leading exponential in the
integrands, so we will have for $n >0 $ (we will use a single index
$n$ because at this order there is no distinction between even and
odd indexes):
\begin{eqnarray}
-4 (2 \sqrt{2} g) \int_{1}^{+\infty} du \, \frac{d^{n}}{du^{n}}
e^{-\frac{g \pi u}{ \sqrt{2} }} = 8 \sqrt{2} g
 \left(-\frac{ \pi g }{  \sqrt{2}} \right)^{n-1} e^{-\frac{\pi g}{ \sqrt{2} }}
+ O\left (e^{-\frac{3 \pi g}{ \sqrt{2} }}\right ) \, . \nonumber
\end{eqnarray}
The case with $n=0$ needs to be treated separately, because we also
have to take into account the contribution of the integral over
$(-\infty,+\infty)$:
\begin{eqnarray}
\int_{-1}^{1} du \, \left[ \frac{\cosh\frac{g \pi u}{ \sqrt{2} }
}{\cosh \sqrt{2} g \pi u} \right] = \int_{-\infty}^{+\infty} du \,
\left[ \frac{\cosh\frac{g \pi u}{ \sqrt{2} } }{\cosh \sqrt{2} g \pi
u} \right] -2 \int_{1}^{+\infty} du \, \left[ \frac{\cosh\frac{g \pi
u}{ \sqrt{2} } }{\cosh \sqrt{2} g \pi u} \right] \nonumber \, .
\end{eqnarray}
We end up with
\begin{eqnarray}
2 (2 \sqrt{2} g) \int_{-1}^{1} du \, \left[ \frac{\cosh\frac{g \pi
u}{ \sqrt{2} } }{\cosh \sqrt{2} g \pi u} \right] = 4 \sqrt{2} - 8
\sqrt{2} g  \left(\frac{ \sqrt{2} }{ g \pi} \right) e^{-\frac{\pi g
}{ \sqrt{2} }} + O\left (e^{-\frac{3 \pi g}{ \sqrt{2} }}\right ) \, .
\end{eqnarray}
We stress that the above integrals have the same structure for any
$n$ only at leading order for large $g$. Next to leading orders will
differ because of the  different form of the integrands.

It is possible to show that, taking the leading exponential only,
the integrals over $|u|>1$ take the form
\begin{eqnarray}
2 \left(-\frac{ \pi g }{  \sqrt{2}} \right)^{n}  \int_{1}^{+\infty}
du \, e^{-\frac{\pi g u}{ \sqrt{2} }} \left[
 \int_0^{+\infty} dt  \, \cos(ut) \,(\Sigma^{(0)}_-(t)  - \Sigma^{(0)}_+( t) )  + \int_0^{+\infty} dt  \,
\sin(ut) \, (\Sigma^{(0)}_-(t)  + \Sigma^{(0)}_+( t) ) \right]   \,
, \nonumber
\end{eqnarray}
which is accurate up to $O\left (e^{-\frac{3 \pi g}{ \sqrt{2} }}\right )$ terms.
The integral was estimated in \cite{BK}, and taking into account the
difference with the notations of that paper\footnote{It is easy to
work out the following relations between our $g$,
$\Sigma_{\pm}^{(0)}$ and the same quantities $g_{BK}$,
$\Gamma_{\pm}^{(0)}$ in the paper \cite{BK}
\begin{equation}
g_{BK} = \frac{g}{\sqrt{2}}, \ \ \ \ \ \Gamma_{\pm}^{(0)} =
\frac{\Sigma_{\pm}^{(0)}}{2 \sqrt{2}g}. \label {bkmap}
\end{equation}
}, we have
\begin{eqnarray}
&& \int_{1}^{+\infty} du \, e^{-\frac{g \pi u}{ \sqrt{2} }} \left[
 \int_0^{+\infty} dt  \, \cos(ut) \,(\Sigma^{(0)}_-(t)  - \Sigma^{(0)}_+( t) )  + \int_0^{+\infty} dt  \,
\sin(ut) \, (\Sigma^{(0)}_-(t)  + \Sigma^{(0)}_+( t) ) \right]   =
\nonumber \\
&& = - \frac{\pi}{ \sqrt{2}} \, m(g) + \frac{8}{\pi}   e^{-\frac{
\pi g}{ \sqrt{2} }} + O\left (e^{- \frac{3 \pi g}{ \sqrt{2} }}\right )\, , \label{BKeq}
\end{eqnarray}
where $m(g)$ is defined as the part of the integral (\ref {BKeq}) proportional to
$e^{-\frac{\pi g}{\sqrt 2}}$ (cf. also \cite{BZJ}):
\begin{equation}
m(g)   =  k \, g^{1/4} e^{-\frac{\pi g}{\sqrt 2}} [1+\sum _{n=1}^{\infty} \frac {k_n}{g^n}], \ \ \ \
\  k= \frac{2^{5/8} \pi^{1/4}}{ \Gamma(5/4)}.
\end{equation}
It is interesting to notice that the $O(1/g)$ in the previous
equation stands for power-like corrections which in can be
computed by inspecting the sub-leading terms of the l.h.s. of
(\ref{BKeq}).

If we put everything together we have,
\begin{eqnarray}
{\mathcal B}_{2k-1}(g) & = &  (-1)^k \left[  \frac {16 \sqrt{2}}{\pi
^2} \left(\frac{ \pi }{{2}} \right)^{2k} e^{-\frac{ \pi g}{ \sqrt{2}
}}- \frac {2 \sqrt{2}}{\pi }\left(\frac{ \pi }{{2}} \right)^{2k}
\left( - \frac{\pi}{ \sqrt{2}} \, m (g) + \frac{8}{\pi}  e^{-\frac{
\pi g}{ \sqrt{2} }} \right) \right] + O(e^{-\frac{3 \pi g
}{ \sqrt{2} }}) \, ,
\nonumber \\
{\mathcal C}_{2 k}(g) & = &  (-1)^k \left[ -4 \sqrt{2}
 \left(\frac{ \pi }{{2}} \right)^{2k-1} e^{-\frac{\pi g}{ \sqrt{2} }}
+{\sqrt {2}}\left(\frac{ \pi }{{2}} \right)^{2k} \left( -
\frac{\pi}{ \sqrt{2}} \, m (g)+ \frac{8}{\pi}  e^{-\frac{ \pi g}{
\sqrt{2} }} \right) \right]  + O(e^{- \frac{3 \pi g
}{ \sqrt{2} }}) \, ,
\nonumber \\
{\mathcal C}_{0}(g) & = &  - \frac {8 \sqrt{2}} {\pi} e^{-\frac{
\pi g}{ \sqrt{2} }}+4  + {\sqrt {2}} \left( - \frac{\pi}{ \sqrt{2}}
\, m (g)+ \frac{8}{\pi}  e^{-\frac{\pi g}{ \sqrt{2} }} \right) + O(e^{- \frac{3\pi g
}{ \sqrt{2} }})
  \, .
\nonumber
\end{eqnarray}
We notice that the first and the last terms always cancel, hence we
have eventually:
\begin{eqnarray}
{\mathcal B}_{2k-1}(g) & = & (-1)^{k}  \left( \frac{\pi}{2}
\right)^{2k} \, 2\, m (g)+ O\left (e^{- \frac{3 \pi g}{ \sqrt{2} }}\right ) \, ,
\nonumber \\
{\mathcal C}_{2 k}(g) & = & (-1)^{k+1}  \left( \frac{\pi}{2}
\right)^{2k} \, \pi \, m (g) + O\left (e^{- \frac{3 \pi g}{ \sqrt{2} }}\right )
\, ,  \ \ \ \ \ \ k>0 \, ,
\nonumber \\
{\mathcal C}_{0}(g) & = &   4 - \pi \, m (g) + O\left (e^{-\frac{3 \pi
g}{ \sqrt{2} }}\right ) = \ \sigma_H^{(0);(0)}. \nonumber
\end{eqnarray}
Therefore, we end up with the following estimates at large $g$:
\begin{eqnarray}
\sqrt{2} g \tilde{S}^{(k)}_1 (g)& = &-\frac{1}{4 \pi} {\mathcal
B}_{2k-1}(g)  =  \frac{(-1)^{k+1}}{4 \pi}  \left( \frac{\pi}{2}
\right)^{2k} \, 2\, m (g) + O\left (e^{- \frac{3 \pi g}{ \sqrt{2} }}\right ) ,
\nonumber \\
\sigma_H^{(0);(2k)} &=&i^{-2k} {\mathcal C}_{2 k}(g)  =  - \left(
\frac{\pi}{2} \right)^{2k} \, \pi \, m(g) + O\left (e^{- \frac{3\pi g}{ \sqrt{2} }}\right )\, ,  \ \ \ \ \ \ k>0 \, ,
\label {cSma} \\
\sigma_H^{(0);(0)} & =& {\mathcal C}_{0}(g)  =    4 - \pi \, m (g)+ O\left (e^{- \frac{3\pi g}{ \sqrt{2} }}\right )
\, . \nonumber
\end{eqnarray}
It is important to point out that, at this level of accuracy, eq.~(\ref{BKeq})
is the same for all the $\mathcal B$, $\mathcal C$.
As a consequence, our estimates (\ref{cSma}), together with
the recursion relations (\ref{recurcoeff}) and (\ref{recurdens}) ensure that $m(g)$ is the same for all the $f_n(g)$, $n>0$.

We conclude with few words on the coefficient relevant for the
computation of $f_1(g)$. Starting from (\ref {intone}) and using the
BKK trasformation (\ref {bkk}), we can obtain
\begin{eqnarray}
{\sqrt {2}}g S_1^{(1)}(g)&=&-\int _{0}^{+\infty}\frac {dt}{2t}\Bigl
[ \frac {\sinh \frac {t}{2 {\sqrt {2}}g}}{\cosh \frac {t}{{\sqrt
{2}}g}}\left (\Sigma _-^{(0)}(t)-
\Sigma _+^{(0)}(t) \right )+ \nonumber \\
&+& \left (1-\frac {\cosh \frac {t}{2 {\sqrt {2}}g}}{\cosh \frac
{t}{{\sqrt {2}}g}}\right ) \left (\Sigma _-^{(0)}(t)+ \Sigma
_+^{(0)}(t) \right ) \Bigr ] \, ,
\end{eqnarray}
which, after the use of the map (\ref {bkmap}), coincides with (53)
of \cite {BK}. Therefore, the strong coupling analysis can be performed along the lines depicted above.

\section{Some symbolic manipulations with {\it Mathematica}$^{\textrm{\textregistered}}$}
\setcounter{equation}{0}

Equations (\ref{f3all} - \ref{f8all}) in the main text can be derived using {\it Mathematica}$^{\textrm{\textregistered}}$ by means of a direct
implementation of equations (\ref{jexp},\ref{recurcoeff}). We begin with some preliminary definitions which are useful for the calculation
{
\footnotesize
\verb"      "\\
\verb"          "\\
\verb"Clear[Combi]"\\
\verb"Combi[k_, l_] := Combi[k, l] = Block[{fin, fin1, enf, t, ind, w, j, en},"\\
\verb"fin = {Table[0, {w, 1, k - l + 1}]}; fin1 = {}; "\\
\verb"Do[Do[Do[j[ind] = fin[[t, ind]], {ind, 1, k - l + 1}];"\\
\verb"Do[{fin = Append[fin, Table[j[f], {f, 1, k - l + 1}]]};"\\
\verb"If[( Sum[j[q], {q, 1, k - l + 1}] == l) && (Sum[q j[q], {q, 1, k - l + 1}] == k), "\\
\verb"fin1 = Append[fin1, Table[j[f], {f, 1, k - l + 1}]]],{j[ind], 0, Min[l, IntegerPart[k/ind]]}],"\\
\verb"{t, 1, Length[fin]}], {ind, 1, k - l + 1}]; en = Union[fin, fin];Union[fin1, fin1]]"\\
\verb"      "\\
\verb"Clear[Prod]     "\\
\verb"Prod[c_, j_, k_, l_] :=Product[(c[p])^j[p]/((j[p])!), {p, 1, k - l + 1}]      "\\
\verb"      "\\
\verb"Clear[xi]      "\\
\verb"xi[n_] := (I)^(n + 1) ((-1)^(n) - 1)/2       "\\
}\\
Then, we introduce equation (\ref{jexp})
{
\footnotesize
\verb"      "\\
\verb"          "\\
\verb"Clear[gamma]"\\
\verb"gamma[k_, l_, n_, c_, h_] := gamma[k, l, n, c, h] = Block[{j, mm},  "\\
\verb"(Sum[(-I)^t Binomial[l-1, t] (((-1)^(t)+1)/2) \[Sigma][n-k, t] h^(l-2-t), {t,0,l-2}])     "\\
\verb"Sum[Clear[j]; Table[{j[p] = Combi[k, l][[ss, p]]}, {p, 1, k - l + 1}];    "\\
\verb"Prod[c, j, k, l], {ss, 1, Length[Combi[k, l]]}] // Expand]    "\\
}\\
and the building block of the recursion relation (\ref{recurcoeff}) for the coefficients $c_n$
{
\footnotesize
\verb"          "\\
\verb"          "\\
\verb"Clear[ConstIter]      "\\
\verb"ConstIter[k_, l_, n_, c_] := ConstIter[k, l, n, c] = Block[{j, mm},"\\
\verb"((I)^(-l+1) \[Sigma][n-k,l-1]) Sum[Clear[j]; Table[{j[p] = Combi[k,l][[ss,p]]}, {p,1,k-l+1}];    "\\
\verb"Prod[c, j, k, l] , {ss, 1, Length[Combi[k, l]]}] // Expand]    "\\
}\\
To compute the generalised scaling functions, we begin with
{
\footnotesize
\verb"          "\\
\verb"          "\\
\verb"\[Sigma][k_, s_] := 0 /; EvenQ[s] == False     "\\
\verb"\[Sigma][2, s_] := 0 "\\
\verb"c[1] = -Pi/(\[Sigma][0, 0]);   "\\
\verb"nmax=8;   "\\
}\\
where the first two lines define some useful properties of the densities $\sigma^{(k),(s)}$, the third
line is the initial condition for the recursion relation (\ref{recurcoeff}), and the last one sets the maximum number of  generalised scaling functions that we want to compute. Then, the coefficients $c_n$ are expressed
in terms of the densities $\sigma^{(k),(s)}$ as follows
{
\footnotesize
\verb"          "\\
\verb"          "\\
\verb"Table[c[nn] = Expand[-(Sum[xi[1] ConstIter[k,1,nn,c], {k,1,nn-1}]  "\\
\verb"+ Sum[ Sum[xi[l] ConstIter[k,l,nn,c], {l,2,k}], {k,1,nn}])/(\[Sigma][0,0])],{nn,2,nmax-2}]; "\\
}\\
and finally, the ratios $\frac{f_n(g)}{2\sqrt{2}g}$ with $n= 3, \dots , nmax$ are obtained by means of
{
\footnotesize
\verb"          "\\
\verb"          "\\
\verb"Table[Print[f[nn] = Sum[Coefficient[Expand[Sum[ Sum[xi[l] gamma[k,l,nn,c,h], {l,1,k}],"\\
\verb"{k, 1, nn}]],h,nt] St[(nt+1)/2], {nt,1,nn,2}]], {nn,3,nmax}];"\\
}\\
whose output can be readily compared with equations (\ref{f3all} - \ref{f8all}) .

\section{Numerics}
\setcounter{equation}{0}

In summary, the main results of this paper may be considered the equations (\ref{f3all} - \ref{f8all}) and the general recursive procedure which has led to them. They provide a compact systematic description of the (first eight) scaling functions $f_n(g)$ at any $g$, not only in the weak, but also in the strong coupling regime. In particular, they are naturally suitable for numerical computations, which will be the topic of this appendix.

As a matter of fact, the building blocks of expressions (\ref{f3all} - \ref{f8all}) are the components
$\tilde S_1^{(r)}(g)$ of the ``reduced systems" (\ref{redSeqn2}), and the densities $\sigma^{(r);(s)}$ (we recall that with this notation we mean the $s$-th derivative of the $r$-th density computed at $u=0$).
Their numerical computation can be achieved with good precision in a broad range of the coupling constant $g$, {\it i.e.} $g \in [0,15]$, allowing us to test the weak and strong coupling regimes, but also giving numerical information about the transition regime.

The numerical technique is that developed in the first reference of \cite{BBKS}.
The aim is to solve the linear systems (\ref{redSeqn2}) for the modes $\tilde S_m^{(r)}(g)$, by truncating the Neumann expansion at a given Bessel (function) order $L$ \footnote{This letter does not mean, in this appendix only, the chain length/angular momentum (cf. Introduction) and thus is a little abuse of notation.}. The structure of the BES kernel (given by the (infinite) matrix $Z_{r,s}(g)$ in the linear system of Bessel functions \cite{BBKS}) is such that the bigger the $L$, the greater the accuracy of the numerical results.

Inspired by results in \cite{BBKS}, we can put down systems (\ref{redSeqn2}) in a matrix form, which is more profitable in a numerical perspective:
\begin{eqnarray}
\label{matsys}
\tilde S^{(r)}_p(g) = b^{(r)}_p(g) - \sum_{m=1}^{\infty} (K^{(m)}_{pm}(g)+2 K^{(c)}_{pm}(g)) \tilde S^{(r)}_m(g) \, ,
\end{eqnarray}
where
\begin{eqnarray}
&& K^{(m)}_{pm}(g) = 2 (N Z)_{pm}, \ \ \ \  K^{(m)}_{pm}(g) = 4 (P N Z Q N Z)_{pm}  \\
&& b^{(r)}(g) = (N + 4 P N Z Q N) \, ({\mathbb I}^{(r)})^{\textrm{T}}  \, ,  \nonumber
\end{eqnarray}
with
\begin{eqnarray}
&&
N= \textrm{diag} (1,2,3, \dots), \ \ \ P=  \textrm{diag} (0,1,0,1 \dots), \ \ \ Q=\textrm{diag} (1,0,1,0 \dots), \nonumber \\
&& {\mathbb I}^{(r)}=({\mathbb I}^{(r)}_1,{\mathbb I}^{(r)}_2,{\mathbb I}^{(r)}_3, \dots) \, .
\end{eqnarray}
The previous equation is remarkably similar to the one coming from the BES equation and numerically solved in \cite{BBKS}: the matrix kernel turns out to be explicitly the same, the only difference being in the forcing term $b^{(r)}_p (g)$ that now involves the integrals $ {\mathbb I}^{(r)}_p$ (\ref{intforterm}).
Of course, the solution can be written as
\begin{eqnarray}
\label{sysnum}
\tilde S^{(r)}(g) \ = \ ({\mathcal I}+K^{(m)}+2 K^{(c)})^{-1}\, b^{(r)}(g) \, ,
\end{eqnarray}
where ${\mathcal I}$ is the identity matrix and, thus, in this way it can be efficiently approximated by truncating the vector solution at length $L$ in a numerical analysis.

From a quantitative point of view, the physically interesting window of
values of the coupling constant is about $g \in [0,15]$, where both the weak and strong coupling regimes can be studied with a satisfactory precision already with a truncation at $L=70$.

This fact implies that the numerical effort is quite small, and  allows the use of {\it Mathematica}$^{\textrm{\textregistered}}$ as the most suitable numerical tool for the solution of the linear problem. As said above, the matrix form (\ref{sysnum}) is particularly easy to be translated in the following {\it Mathematica}$^{\textrm{\textregistered}}$ code:
{
\footnotesize
\verb"          "\\
\verb"          "\\
\verb" II[L_] := II[L] = IdentityMatrix[L]"\\
\verb" NN[L_] := NN[L] = DiagonalMatrix[Table[i, {i, 1, L}]];"\\
\verb" QQ[L_] := QQ[L] = DiagonalMatrix[Table[(1-(-1)^i)/2, {i, 1, L}]];"\\
\verb" PP[L_] := PP[L] = DiagonalMatrix[Table[(1+(-1)^i)/2, {i, 1, L}]];"\\
\verb" Km[L_,g_] := Km[L,g] = 2 NN[L].ZZ[L,g];" \\
\verb" Kc[L_,g_] := Kc[L,g] = 4 PP[L].NN[L].ZZ[L, g].QQ[L].NN[L].ZZ[L,g];"  \\
\verb" b[r_,L_,g_] := b[r,L,g] = (NN[L] + 4 PP[L].NN[L].ZZ[L, g].QQ[L].NN[L]).Integ[r,L,g];" \\
\verb" Mat[L_,g_] := Mat[L, g] = II[L] + Km[L,g] + 2 Kc[L,g];" \\
\verb" InvMat[L_,g_] := InvMat[L,g] = Inverse[Mat[L,g]];" \\
\verb" tS[L_,g_] := tS[L,g] = Inverse[Mat[L,g]].b[r,L,g];" \\
}\\
The only external input needed is the evaluation of $Z$ and $ {\mathbb I}^{(r)}$ (encoded in the arrays \verb"ZZ[L,g]" and  \verb"Integ[r,L, g]"), whose entries  are defined as integrals in (\ref{Zetaint}) and  (\ref{intforterm}), respectively. Even though these low values of $L$ and $g$ would permit to deal with a numerical integration under {\it Mathematica}$^{\textrm{\textregistered}}$, it is far more efficient to use a standard numerical integrator programmed in a C language code. The output of this program is then stored once forever in arrays that can be loaded in a {\it Mathematica}$^{\textrm{\textregistered}}$ notebook when necessary. Then, the solution of the truncated linear system is stored in the table \verb"NeumannModes", where we have one array \verb"tS[r,L,g]" for each value of $g$.

The procedure described before gives the core of the numerical computation.
At this stage, we are in the position to extract the numerical estimates for $\tilde S_1^{(r)}(g)$, and the densities $\sigma_H^{(r);(s)}$. The former is just the first component in the numerical array above for the solution $\tilde S^{(r)}(g)$. According to (\ref{signs}), the latter is given by an infinite sum over the components of  $S^{(r)}(g)$, similar to (\ref{higher}), but now the sum is truncated at $L$, and the integrals over the Bessel functions are computed numerically through a C language code\footnote{As discussed in the main body of the text, this procedure is strictly true for $\sigma_H^{(r);(s)}$ with $r=1,2$, and can be used to obtain the initial values for the recursive relation (\ref{recurdens}) which give $\sigma_H^{(r);(s)}$ with $r>2$.}.

In general, the procedure described above allows us to produce a numerical estimate of the scaling functions $f_n(g)$, and the densities $\sigma^{(r);(s)}$ as functions of the coupling constant in a given range. As an example of the application of the method, we provide some results
for the scaling function $f_3(g)$, and the densities $\sigma^{(0);(0)}$. Since the strong coupling regime is of particular interest, we can focus on the numerical analysis concerning it (this was done for the first time in \cite{FGR}, where we have obtained the mass gap of the $O(6)$ NLSM from the first scaling function $f_1(g)$). The simplest way to achieve reliable quantitative information is the use of the well known best fit procedure based on the ``$\chi^2$" statistical test.
We also recall that
\begin{equation}
f_3(g) =  \frac {\pi ^2}{6 \, [-4+\sigma^{(0);(0)}_H(0)]^2} f_3^{red}(g)
\,  \label {rednonred}
\end{equation}
and hence we will study initially the reduced version $f_3^{red}(g)$. As we already know that the functional form we have to use is exponential, we begin with the following hypothesis for the strong coupling behaviour of $f_3^{red}(g)$, and $\sigma_H^{(0);(0)}$:
\begin{equation}
\sigma^{(0);(0)}_H(0)|_{fit}  = 4 + d_0^{fit} g^{1/4} e^{-\frac{\pi}{\sqrt 2}g}
\, ,
\end{equation}
\begin{equation}
f_3^{red}(g)|_{fit} = c_3^{fit} g^{1/4} e^{-\frac{\pi}{\sqrt 2}g} \, ,
\end{equation}
where the constants $d_0^{fit} $, $c_3^{fit}$ will be fixed by the best fit procedure.
The latter has proved to work remarkably well giving the following estimates
\begin{equation}
d_0^{fit} = -7.1166 \pm 0.0005,
\end{equation}
\begin{equation}
c_3^{fit} = 5.5896 \pm 0.0005 \, ,
\end{equation}
with a $\chi^2 \sim 1$ in the range $g \in [3,12]$, and also a very good degree of
accuracy with respect to the exact estimates (see section 5).

Following this same workflow it is possible to reproduce the analysis of subsection 5.2.



\begin{thebibliography}{xx}


\bibitem{MWGKP}
J.M. Maldacena, {\sl The large N limit of superconformal field
theories and supergravity}, Adv. Theor. Math. Phys.
{\bf 2} (1998) 231 and hep-th/9711200 $\bullet $
S.S. Gubser, I.R. Klebanov, A.M. Polyakov, {\sl Gauge theory
correlators from non-critical string theory},
Phys.Lett. {\bf B428} (1998) 105 and hep-th/9802109 $\bullet $
E. Witten, {\sl Anti-de Sitter space and holography}, Adv. Theor.
Math. Phys. {\bf 2} (1998) 253 and hep-th/9802150;

\bibitem{GKPII}
S.~S.~Gubser, I.~R.~Klebanov and A.~M.~Polyakov,
  Nucl.\ Phys.\  B {\bf 636} (2002) 99
  [arXiv:hep-th/0204051];

\bibitem{FT}
S.~Frolov and A.~A.~Tseytlin,
  JHEP {\bf 0206} (2002) 007
  [arXiv:hep-th/0204226];

\bibitem{LIP}
L.N. Lipatov, {\sl Evolution equations in QCD}, in ``Perspectives in
Hadron Physics'', Prooceedings of the Conference, ICTP, Trieste,
Italy, May 1997, World Scientific (Singapore, 1998);
\bibitem{BDM}
V.M. Braun, S.E. Derkachov, A.N. Manashov, {\sl Integrability of
three particle evolution equations in QCD}, Phys. Rev. Lett. {\bf
81} (1998) 2020 and
hep-ph/9805225 $\bullet $
V.M. Braun, S.E. Derkachov, G.P. Korchemsky, A.N. Manashov, {\sl Baryon distribution amplitudes in QCD}, Nucl. Phys. {\bf B553} (1999) 355 and hep-ph/9902375 $\bullet $
A.V. Belitsky, {\sl Fine structure of spectrum of twist-three operators in QCD}, Phys. Lett. {\bf B453} (1999) 59 and hep-ph/9902361 $\bullet $
A.V. Belitsky, {\sl Renormalization of twist-three operators and
integrable lattice models},  Nucl. Phys. {\bf B574} (2000) 407 and
hep-ph/9907420;
\bibitem{MZ}
J.A. Minahan, K. Zarembo, {\sl The Bethe Ansatz for ${\cal N}=4$
Super Yang-Mills}, JHEP{\bf 03} (2003) 013 and hep-th/0212208;
\bibitem{BS}
N. Beisert, M. Staudacher, {\sl The ${\cal N}=4$ SYM integrable
super spin
chain}, Nucl. Phys. {\bf B670} (2003) 439 and hep-th/0307042 $\bullet $
M. Staudacher, {\sl The factorized S-matrix of CFT/AdS},
JHEP{\bf 05} (2005) 054 and hep-th/0412188 $\bullet $
N. Beisert, M. Staudacher, {\sl Long-range $PSU(2,2|4)$ Bethe Ansatz
for gauge theory and strings}, Nucl. Phys. {\bf B727} (2005) 1 and
hep-th/0504190 $\bullet $
G. Arutyunov, S. Frolov, M. Staudacher, {\sl  Bethe ansatz for
quantum strings}, JHEP{\bf 10} (2004) 016
and hep-th/0406256;
\bibitem{BES}
B. Eden, M. Staudacher, {\sl Integrability and transcendentality},
J.Stat.Mech. {\bf 11} (2006) P014 and hep-th/0603157 $\bullet $
N. Beisert, B.Eden, M. Staudacher, {\sl Transcendentality and
crossing}, J.Stat.Mech.{\bf 07} (2007) P01021 and hep-th/0610251;
\bibitem{FTT}
S. Frolov, A. Tirziu, A.A. Tseytlin, {\sl Logarithmic corrections to
higher twist scaling at strong coupling from AdS/CFT}, Nucl. Phys.
{\bf B766} (2007) 232 and hep-th/0611269;
\bibitem{BGK}
A.V.Belitsky, A.S. Gorsky, G.P. Korchemsky, {\sl Logarithmic scaling
in gauge/string correspondence}, Nucl. Phys. {\bf B748} (2006) 24
and hep-th/0601112;
\bibitem{AM}
L.F. Alday, J.M. Maldacena, {\sl Comments on operators with large
spin}, JHEP{\bf 11} (2007) 019 and arXiv:0708.0672 [hep-th];

\bibitem{FMQR}
D. Fioravanti, A. Mariottini, E. Quattrini, F. Ravanini, {\sl
Excited state Destri-de Vega equation for sine-Gordon and restricted
sine-Gordon models}, Phys. Lett. {\bf B390} (1997) 243
and hep-th/9608091 $\bullet $
C. Destri, H.J. de Vega, {\sl Non linear integral equation and
excited states scaling functions in the sine-Gordon model}, Nucl.
Phys, {\bf B504} (1997) 621 and hep-th/9701107;



\bibitem{FRS}
L. Freyhult, A. Rej, M. Staudacher, {\sl A Generalized Scaling
Function for AdS/CFT}, J. Stat. Mech. (2008) P07015 and
arXiv:0712.2743 [hep-th];



\bibitem{BBKS}
M. K. Benna, S. Benvenuti, I. R. Klebanov, A. Scardicchio, {\sl A
Test of the AdS/CFT Correspondence Using High-Spin Operators}, Phys.
Rev. Lett. {\bf 98} (2007) 131603 and
hep-th/0611135 $\bullet $
L. F. Alday, G. Arutyunov, M. K. Benna, B. Eden, I. R. Klebanov,
{\sl On the Strong Coupling Scaling Dimension of High Spin
Operators}, JHEP{\bf 04} (2007) 082 and hep-th/0702028 $\bullet $
I. Kostov, D. Serban and D. Volin, {\sl Strong coupling limit of
Bethe Ansatz equations}, Nucl. Phys. {\bf B789} (2008) 413 and
hep-th/0703031 $\bullet $
M. Beccaria, G.F. De Angelis, V. Forini, {\sl  The Scaling function
at strong coupling from the quantum string Bethe equations},
JHEP{\bf 04} (2007) 066 and hep-th/0703131;
\bibitem{BKK}
B. Basso, G.P. Korchemsky, J. Kotanski, {\sl Cusp anomalous
dimension in maximally supersymmetric Yang-Mills theory at strong
coupling}, Phys. Rev. Lett. {\bf 100} (2008) 091601 and
arXiv:0708.3933 [hep-th];
\bibitem{KSV}
I.~Kostov, D.~Serban and D.~Volin, {\sl Functional BES equation},
arXiv:0801.2542 [hep-th];
\bibitem{CK}
P.Y. Casteill, C. Kristjansen, {\sl The strong coupling limit of the scaling function from the quantum string Bethe ansatz}, Nucl. Phys. {\bf B785} (2007) 1 and arXiv:0705.0890 [hep-th] $\bullet $
A.V. Belitsky, {\sl Strong coupling expansion of Baxter equation in ${\cal N}=4$ SYM}, Phys. Lett. {\bf B659} (2008) 732 and arXiv:0710.2294 [hep-th] $\bullet $
R. Roiban, A.A. Tseytlin, {\sl Spinning superstrings at two loops: strong coupling corrections to dimensions of large-twist SYM operators}, Phys. Rev. {\bf D77}(2008) 066006 and arXiv:0712.2479 [hep-th] $\bullet $
N. Gromov, {\sl Generalized scaling function at strong coupling},
arXiv:0805.4615 [hep-th];

\bibitem{BFR2}
D. Bombardelli, D. Fioravanti, M. Rossi, {\sl Large spin corrections
in ${\cal N}=4$ SYM $sl(2)$: still a linear integral equation}, Nucl. Phys. {\bf B810} (2009) 460 and arXiv:0802.0027 [hep-th];
\bibitem{FGR}
D. Fioravanti, P. Grinza, M. Rossi, {\sl Strong coupling for planar
 ${\cal N}=4$ SYM: an all-order result}, Nucl.
Phys. {\bf B810} (2009) 563 and arXiv:0804.2893 [hep-th];
\bibitem{BK}
B. Basso, G.P. Korchemsky, {\sl Embedding nonlinear $O(6)$ sigma
model into
 ${\cal N}=4$ super-Yang-Mills theory}, arXiv:0805.4194 [hep-th];
\bibitem{FGR2}
D. Fioravanti, P. Grinza, M. Rossi,  {\sl The generalised scaling
function: a note}, arXiv:0805.4407 [hep-th];
\bibitem{BF}
F. Buccheri, D. Fioravanti, {\sl The integrable $O(6)$ model and the
correspondence: checks and predictions}, arXiv:0805.4410 [hep-th];
\bibitem{KL}
A.V. Kotikov, L.N. Lipatov, {\sl On the highest transcendentality in
 ${\cal N}=4$ SUSY}, Nucl. Phys. {\bf B769} (2007) 217 and hep-th/0611204;



\bibitem{NOI}
G. Feverati, D. Fioravanti, P. Grinza, M. Rossi, {\sl On the finite
size corrections of anti-ferromagnetic anomalous dimensions
in ${\cal N}=4$ SYM}, JHEP{\bf 05} (2006) 068 and hep-th/0602189 $\bullet $
G. Feverati, D. Fioravanti, P. Grinza, M. Rossi, {\sl Hubbard's
Adventures in ${\cal N}=4$ SYM-land? Some non-perturbative
considerations on finite length operators},  J.Stat.Mech. {\bf 02}
(2007) P001 and hep-th/0611186 $\bullet $
D. Fioravanti, M. Rossi, {\sl On the commuting charges for the
highest dimension $SU(2)$ operators in planar ${\cal N}=4$ SYM},
JHEP{\bf 08} (2007) 089 and
arXiv:0706.3936 [hep-th] $\bullet $
D. Bombardelli, D. Fioravanti, M. Rossi, {\sl  Non-linear integral
equations in N = 4 SYM}, Proceedings of
International Workshop on Recent Advances in Quantum Integrable
Systems (RAQIS 07), Annecy-le-Vieux, France, 11-14 Sep 2007 and
arXiv:0711.2934 [hep-th];

\bibitem{faa}
R.P. Stanley, S. Fomin, {\sl Enumerative combinatorics}, Vol. 2,
Cambridge Studies in Advanced Mathematics 62;
\bibitem {BEC}
M. Beccaria, {\sl The generalized scaling function of AdS/CFT and
semiclassical string theory}, arXiv:0806.3704 [hep-th];

\bibitem{ABJM}
O. Aharony, O. Bergman, D.L. Jafferis, J. Maldacena, {\sl ${\cal
N}=6$ superconformal Chern-Simons matter theories, $M$-$2$ branes
and their gravity duals},
arXiv:0806.1218 [hep-th] $\bullet $
M. Benna, I. Klebanov, T. Klose, M. Smedback, {\sl Superconformal
Chern-Simons Theories and $AdS_4/CFT_3$ correspondence},
arXiv:0806.1519 [hep-th];
\bibitem{MZ2}
J.A. Minahan, K. Zarembo, {\sl The Bethe ansatz for superconformal
Chern-Simons}, arXiv:0806.3951 [hep-th];
\bibitem{GSY}
D. Bak, S.J. Rey, {\sl Integrable spin chain in superconformal
Chern-Simons theory}, arXiv:0807.2063 [hep-th];
\bibitem{AF}
G. Arutyunov, S. Frolov, {\sl Superstrings on $AdS_4/CFT_3$ as a Coset sigma-model}, arXiv:0806.4940 [hep-th]
$\bullet $ B. Stefanski jr., {\sl Green-Schwarz action for type IIA strings on
$\text{AdS}_4\times\text{CP}^3$}, arXiv:0806.4948 [hep-th]  $\bullet $ G. Grignani, T. Harmark, M. Orselli,
{\sl The $SU(2) \times SU(2)$ sector in the string dual of $N=6$ superconformal Chern-Simons theory},
Nucl. Phys. {\bf B810} (2009) 115 and arXiv:0806.4959 [hep-th]; $\bullet $ D.~Astolfi, V.~G.~M.~Puletti, G.~Grignani {\it et al.}, ``Finite-size corrections in the SU(2) x SU(2) sector of type IIA string theory on AdS(4) x CP**3,'' Nucl.\ Phys.\  {\bf B810 } (2009)  150-173 [arXiv:0807.1527 [hep-th]].
\bibitem{GV}
N. Gromov, P. Vieira, {\sl The all loop $AdS_4/CFT_3$ Bethe Ansatz},
arXiv:0807.0777 [hep-th];
\bibitem{AVGHO}
C. Ahn, R.I. Nepomechie, {\sl  ${\cal N}=6$ super Chern-Simons theory S-matrix and all-loop Bethe Ansatz equations}, arXiv:0807.1924 [hep-th] $\bullet $
T. McLoughlin, R. Roiban, {\sl Spinning strings at one-loop in
 $\text{AdS}_4\times\text{P}^3$}, arXiv:0807.3965 [hep-th] $\bullet $
L.F. Alday, G. Arutyunov, D. Bykov, {\sl Semiclassical quantization of spinning strings in $\text{AdS}_4\times\text{CP}^3$}, arXiv:0807.4400 [hep-th] $\bullet $
C. Krishnan, {\sl  $AdS_4/CFT_3$ at one loop}, arXiv:0807.4561 [hep-th] $\bullet $
N. Gromov, V. Mikhaylov, {\sl Comment on scaling function in
$\text{AdS}_4\times\text{CP}^3$}, arXiv:0807.4897 [hep-th];
\bibitem{BZJ}
E. Brezin, J. Zinn-Justin, {\sl Spontaneous breakdown of continuous
symmetries near two dimensions}, Phys. Rev. {\bf B14} (1976) 3110.






\end{thebibliography}
\end{document}